\documentclass[fleqn,usenatbib]{mnras}
\usepackage{CJKutf8}

\usepackage{bm}

\usepackage[T1]{fontenc}
\usepackage{longtable}
\usepackage{tabu}
\usepackage{caption}
\usepackage{xcolor,colortbl}

\DeclareRobustCommand{\VAN}[3]{#2}
\let\VANthebibliography\thebibliography
\def\thebibliography{\DeclareRobustCommand{\VAN}[3]{##3}\VANthebibliography}

\usepackage{graphicx}	% Including figure files
\usepackage{amsmath}	% Advanced maths commands
\usepackage{amssymb}	% Extra maths symbols
\usepackage{float}
\usepackage{newtxtext,newtxmath}
\usepackage{soul}
\title[Searching for extra-tidal stars]{Searching for the extra-tidal stars of globular clusters using high-dimensional analysis and a core particle spray code}

\author[Grondin et al.]{
Steffani M. Grondin,$^{1}$\thanks{E-mail: steffani.grondin@astro.utoronto.ca}
Jeremy J. Webb,$^{1}$
Nathan W. C. Leigh,$^{2,3}$
Joshua S. Speagle (\begin{CJK*}{UTF8}{gbsn}沈佳士\ignorespacesafterend\end{CJK*}),$^{5, 1, 4}$
\and 
and Reem J. Khalifeh$^{1}$
\\
$^{1}$ David A. Dunlap Department of Astronomy \& Astrophysics, University of Toronto
50 St. George St, Toronto M5S 3H4, Canada \\
$^{2}$ Departamento de Astronom\'ia, Facultad Ciencias F\'isicas y Matem\'aticas, Universidad de Concepci\'on, Av. Esteban Iturra s/n Barrio Universitario,\\ Casilla 160-C, Concepci\'on, Chile \\
$^{3}$ Department of Astrophysics, American Museum of Natural History, New York, NY 10024, USA \\
$^{4}$ Dunlap Institute for Astronomy and Astrophysics, University of Toronto, 50 St George Street, Toronto, ON M5S 3H4, Canada \\
$^{5}$ Department of Statistical Sciences, University of Toronto, 100 St George St, Toronto, ON M5S 3G3, Canada \\
}

\date{Accepted XXX. Received YYY; in original form ZZZ}

\pubyear{2021}

\begin{document}
\label{firstpage}
\pagerange{\pageref{firstpage}--\pageref{lastpage}}
\maketitle

\begin{abstract}
Three-body interactions can eject stars from the core of a globular cluster, causing them to enter the Galactic halo as extra-tidal stars. While finding extra-tidal stars is imperative for understanding cluster evolution, connecting isolated extra-tidal field stars back to their birth cluster is extremely difficult. In this work, we present a new methodology consisting of high-dimensional data analysis and a particle spray code to identify extra-tidal stars of \textit{any} Galactic globular cluster using M3 as a case study. Using the t-Stochastic Neighbour Embedding (t-SNE) and Uniform Manifold Approximation and Projection (UMAP) machine learning dimensionality reduction algorithms, we first identify a set of 103 extra-tidal candidates in the APOGEE DR17 data catalogue with chemical abundances similar to M3 stars. To confirm each candidate's extra-tidal nature, we introduce \texttt{corespray} — a new \texttt{Python}-based three-body particle spray code that simulates extra-tidal stars for any Galactic globular cluster. Using Gaia EDR3 proper motions and APOGEE DR17 radial velocities, we apply multivariate Gaussian modelling and an extreme deconvolution to identify the extra-tidal candidates that are more likely to be associated with a distribution of \texttt{corespray}-simulated M3 extra-tidal stars than the field.  Through these methods, we identify 10 new high-probability extra-tidal stars produced via three-body interactions in M3. We also explore whether any of our extra-tidal candidates are consistent with being ejected from M3 through different dynamical processes. Future applications of \texttt{corespray} will yield better understandings of core dynamics, star formation histories and binary fractions in globular clusters. 
\end{abstract}


%%%%%%%%%%%%%%%%%%%%%%%%%%%%%%%%%%%%%%%%%%%%%%%%%%

%%%%%%%%%%%%%%%%% BODY OF PAPER %%%%%%%%%%%%%%%%%%

\section{Introduction}\label{sec:intro}

Globular clusters (GCs) are some of the most ancient structures in the Milky Way. These collections of stars have been measured to have ages $\gtrsim 12.5$ Gyr, providing evidence that many GCs likely formed around the time of cosmic reionization \citep{2018RSPSA.47470616F}. As such, it is believed that the formation of most GCs occurred before star formation in galaxies \citep{2017MNRAS.469L..63R} and thus likely played a crucial role in forming the Galaxy \citep{2019A&ARv..27....8G}. While GCs are useful for learning about galaxy formation, their dense nature also makes them ideal systems to probe how galaxies evolve over time. Specifically, GC evolution informs star formation histories from initial GC size and mass functions and constrains binary fractions from GC population synthesis studies \citep{2010MNRAS.406.2000M, 2005MNRAS.358..572I}. Furthermore, since GCs are densely packed groups of hundreds of thousands to millions of stars with half-light radii $\leq 10$ pc, they are ideal environments for learning how dynamical processes influence stars over time \citep{2019A&ARv..27....8G, 2008MNRAS.385L..20V}. Importantly, while stellar interactions like two-body relaxation and tidal stripping are primarily responsible for mass loss of a GC (and can thus lead to stellar streams or tidal tails), they do not probe a GC's core or binary systems. Thus, to investigate how stars and binaries escape the \textit{core} of a GC, \textit{three-body interactions} are imperative.

While massive star evolution drives internal GC evolution in the first $\sim 1$ Gyr after formation, GCs generally evolve due to gravitational forces from the host galaxy \citep{2018ApJ...861...69C} and gravitational encounters between cluster stars \citep{2003gmbp.book.....H}. An important example of the latter occurs in a three-body interaction \citep{leigh13}, where the high number density and low volume in a GC's core allow for a close encounter between three stars. \footnote{Note that three-body interactions are not the same as triple systems, which are gravitationally bound systems that can become unstable. \cite{2011MNRAS.416.1410L} showed that the presence of triple systems in GCs is rare and as such, they are not considered to be the precursor dynamical interactions yielding extra-tidal stars in this study.} Assuming all stars are point-particles (i.e. mergers cannot occur), this 1+2 reaction results in the ejection of a single star from the three-body system while causing the remaining two stars to form a binary \citep[e.g.][]{stoneleigh19,manwadkar20,manwadkar21}. The ejection of the kicked star will cause the binary to recoil, causing it to get kicked in the opposite direction \citep{3body}. In some cases, these kicks can result in stars moving beyond a cluster's tidal radius.

Defined both theoretically and observationally \citep[e.g.][]{1962AJ.....67..471K, 1983AJ.....88..338I, 1992ApJ...386..506O, 2013ApJ...764..124W}, the tidal radius of a cluster represents the limit where gravitational forces from the host galaxy balance gravitational forces from the GC. However, a GC's tidal radius is not constant, rather it changes as a function of time along the cluster's orbit through the Galaxy \citep{2011MNRAS.418..759R, 2013ApJ...764..124W}. Specifically, the GC's tidal radius will be at a minimum when the cluster is at perigalacticon and a maximum when the cluster is at apogalacticon. \footnote{ A cluster's tidal radius at apogalacticon can be computed by applying a correction to the tidal radius at perigalacticon. In this study, we perform the transformation via Equation 8 in \cite{2013ApJ...764..124W}.} Although it takes time for stars to completely leave the cluster after becoming energetically unbound \citep{1987ApJ...322..123L}, cluster stars that surpass the tidal radius will eventually no longer remain gravitationally bound to the parent GC \citep{2018ApJ...861...69C}. Thus, if a star receives enough energy from a kick during a three-body interaction, it will migrate out of the GC's core and enter the Galactic halo as an \textit{extra-tidal star}.

Dense GCs undergo higher numbers of three-body encounters than their less dense counterparts, which leads to more stars escaping the cluster \citep[e.g.][]{leigh11,leigh12,leigh15,leigh17,leigh18, barrera21,parischewsky21,reinoso22}. Generally, these types of three-body interactions will occur approximately once every $\sim 10$ Myr within GCs \citep{leigh11}. Since these encounters happen while a GC orbits the Galaxy, a given cluster's extra-tidal stars can be dispersed all throughout the Milky Way. Thus, connecting extra-tidal stars back to their parent cluster is difficult, especially when the stars are isolated and located far away. However, all stars in a GC form from either the same giant molecular cloud or GC mergers and as such, the chemical abundances of stars originating in a specific GC should be similar \citep{2006AJ....131..455D, 2007AJ....133.1161D, 2016ApJ...817...49B, 2018MNRAS.475.1410P}. Since certain kinematic quantities are conserved along the GC's orbit, stars that receive a velocity kick due to a three-body interaction should have similar kinematics as the cluster itself, no matter where along the cluster’s orbit it escaped \citep{2008gady.book.....B}. Thus, finding a star beyond the tidal radius of a GC that exhibits similarities in both (i) chemical abundances and (ii) conserved kinematic orbital properties would be strong evidence to link the extra-tidal star back to its original birth cluster.

 Clusters that are especially dense and contain many members are ideal environments to search for new extra-tidal stars. As presented in \cite{2020MNRAS.492.1641M}, one cluster meeting both of the aforementioned requirements is M3 (NGC 5272). M3 has a right ascension of $\alpha=205.548^{\circ}$, declination of $\delta = 28.377^{\circ}$, proper motion of $(\mu_{\alpha}\cos{\delta}, \mu_{\delta}) = (-0.142, -2.647)$ km/s, radial velocity of $v_{r} = -147.28$km/s and a heliocentric distance of 10.2kpc \citep{Vasiliev2019} \footnote{Orbital parameters of Galactic GCs can be accessed via the \cite{Vasiliev2019} cluster catalogue.}. At apogalacticon, M3 has a tidal radius of $r_{t} = 159.03$pc and an angular size of $r_{t} \sim 0.895^{\circ}$ \citep{2018MNRAS.478.1520B}\footnote{Structural parameters of Galactic GCs can be accessed via the online database at \href{https://people.smp.uq.edu.au/HolgerBaumgardt/globular/}{https://people.smp.uq.edu.au/HolgerBaumgardt/globular/.}}. 
 
 Although \cite{2000A&A...359..907L} found extra-tidal structure around M3 through the identification of a tidal tail, later studies by \cite{2006ApJ...639L..17G}, \cite{2010A&A...522A..71J} and \cite{2014MNRAS.445.2971C} found no evidence for either extra-tidal stars or structure altogether. However, using the first data release from LAMOST, \cite{2016ApJ...829..123N} identified extra-tidal stars around M3 (and M13). The authors first searched for cluster members of M3 by selecting stars that were within a $5^{\circ}$ radius around the GC's centre and had radial velocities within $\pm 2\sigma$ of M3's radial velocity. Through establishing $V$ versus $V-K$ and $\log(T_{\text{eff}})$ versus $\log(g)$ limits and only accepting stars within $10$ mas $yr^{-1}$ of the GC's proper motion, \cite{2016ApJ...829..123N} identified eight new extra-tidal stars of M3.
 
 Since the \cite{2016ApJ...829..123N} study, advances in machine learning have introduced new methods for identifying similar stars in high-dimensional parameter spaces. Recent work by \cite{2020ApJ...900..146C} employed the use of both chemical tagging and kinematic analyses to search for extra-tidal stars around M53 (NGC 5024) and NGC 5053 — two metal-poor GCs in the Galactic halo. Searching a $20^{\circ} \times 10^{\circ}$ field around the clusters, \cite{2020ApJ...900..146C} used abundances and radial velocities from the 14th data release (DR14) of the Apache Point Observatory Galactic Evolution Experiment (APOGEE). Using this data, the authors applied a t-Stochastic Neighbour Embedding (t-SNE) machine learning algorithm to identify stars that were both chemically and kinematically similar, allowing them to define members and extra-tidal stars of each cluster. Combined with additional metallicity and radial velocity examinations for stars in which t-SNE was not applicable, \cite{2020ApJ...900..146C} identified three and four extra-tidal stars for M53 and NGC 5053 respectively. Chemical tagging and kinematic constraints have even been used to identify stars that formed in the same birth cluster, despite the host cluster having already dissolved \citep{2020MNRAS.494.2268W, 2020MNRAS.496.5101P}. 

Increasing the known sample of extra-tidal stars in our Galaxy will allow us to better constrain the underlying Galactic gravitational potential, possible overdensities and binary fractions, ultimately allowing for advancements in our understanding of the evolution of both Galactic GCs and the Milky Way itself. Thus, in this study we build off the work of both \cite{2016ApJ...829..123N} and \cite{2020ApJ...900..146C} to present a new methodology for identifying extra-tidal stars in and around \textit{any} Galactic GC, using M3 as a case study. In Section \ref{sec:observational}, we outline an observational identification scheme of extra-tidal candidates using stars in the 17th data release (DR17) of APOGEE. Specifically, we present methodology to identify stars that are chemically similar to suspected parent GCs using two different machine learning clustering algorithms — t-SNE \citep[as used in][]{2020ApJ...900..146C} and Uniform Manifold Approximation and Projection (UMAP). To further confirm the extra-tidal nature of the observationally-identified extra-tidal candidates, Section \ref{sec:theoretical} presents \texttt{corespray} — a new Python-based code that uses three-body dynamics to simulate the creation of extra-tidal stars and their corresponding recoil binaries. \texttt{Corespray} allows the user to explore where extra-tidal stars of a given GC could end up in a variety of different parameter spaces by taking into consideration the individual GC's mass, tidal radius, central velocity dispersion, central escape velocity, central potential and core density. Section \ref{sec:newstars} combines our observational identification and theoretical confirmation methods to identify new high-probability extra-tidal stars of M3. We discuss our results in Section \ref{sec:discussion} and conclude in Section \ref{sec:conclusions}. 

\section{Data}
\label{sec:data}

To identify extra-tidal candidates, we search the DR17 catalogue from the APOGEE survey \citep{2022ApJS..259...35A}. APOGEE is a high-resolution ($R \sim 22,500$), high signal-to-noise ratio (SNR $>100$), infrared ($1.51-1.70 \mu m$) spectroscopic survey \citep{2017AJ....154...94M}. As our observational extra-tidal identification scheme centres around chemical abundances, we utilize the high-quality \texttt{astroNN} abundances derived by \cite{2019MNRAS.483.3255L, 2019MNRAS.489.2079L} and \cite{2019MNRAS.490.4740B}. \texttt{AstroNN} abundances are determined through a neural network using APOGEE data, providing us with 19 chemical abundances and stellar parameters for an initial sample of 694,932 stars. To examine the kinematics of extra-tidal star candidates throughout our analyses, we use Gaia Early Data Release 3 (EDR3) proper motions \citep{2021A&A...649A...1G} and APOGEE DR17 barycentric radial velocities \citep{2022ApJS..259...35A}. Of our initial sample, 685,978 stars have Gaia EDR3 proper motions. 

To begin our search for extra-tidal candidates, we first follow the work of \cite{2020ApJ...900..146C} and select APOGEE DR17 stars that are within a $10^{\circ} \times 10^{\circ}$ field of view (FOV) around the GC's centre. This limited FOV restricts our search to stars that have recently escaped the GC and will have had their orbits minimally affected by Galactic substructure. For M3, this spatial cut reduces our sample to 5240 stars, just under $1\%$ of the \texttt{astroNN} APOGEE DR17 catalogue. Furthermore, 187 of these stars do not have any reported chemical abundances, reducing the data set to 5053 stars. Often, APOGEE DR17 contains multiple spectra for the same source. To filter out duplicate sources, we select stars that either (i) only contain a single measurement (1918 stars) or (ii) have the highest signal-to-noise (SNR) spectrum of a source with multiple measurements (1391 stars). Removing duplicate sources reduces our sample size to 3309 stars. Finally, to ensure that we only include stars with high quality spectra in our search, we impose an additional SNR cut, only keeping stars with SNR $\geq 50$ (which mostly yields red giant stars). Through this filtering, our M3 data set contains 3212 unique stars. 

\begin{figure}
    \centering
    \includegraphics[width=\columnwidth]{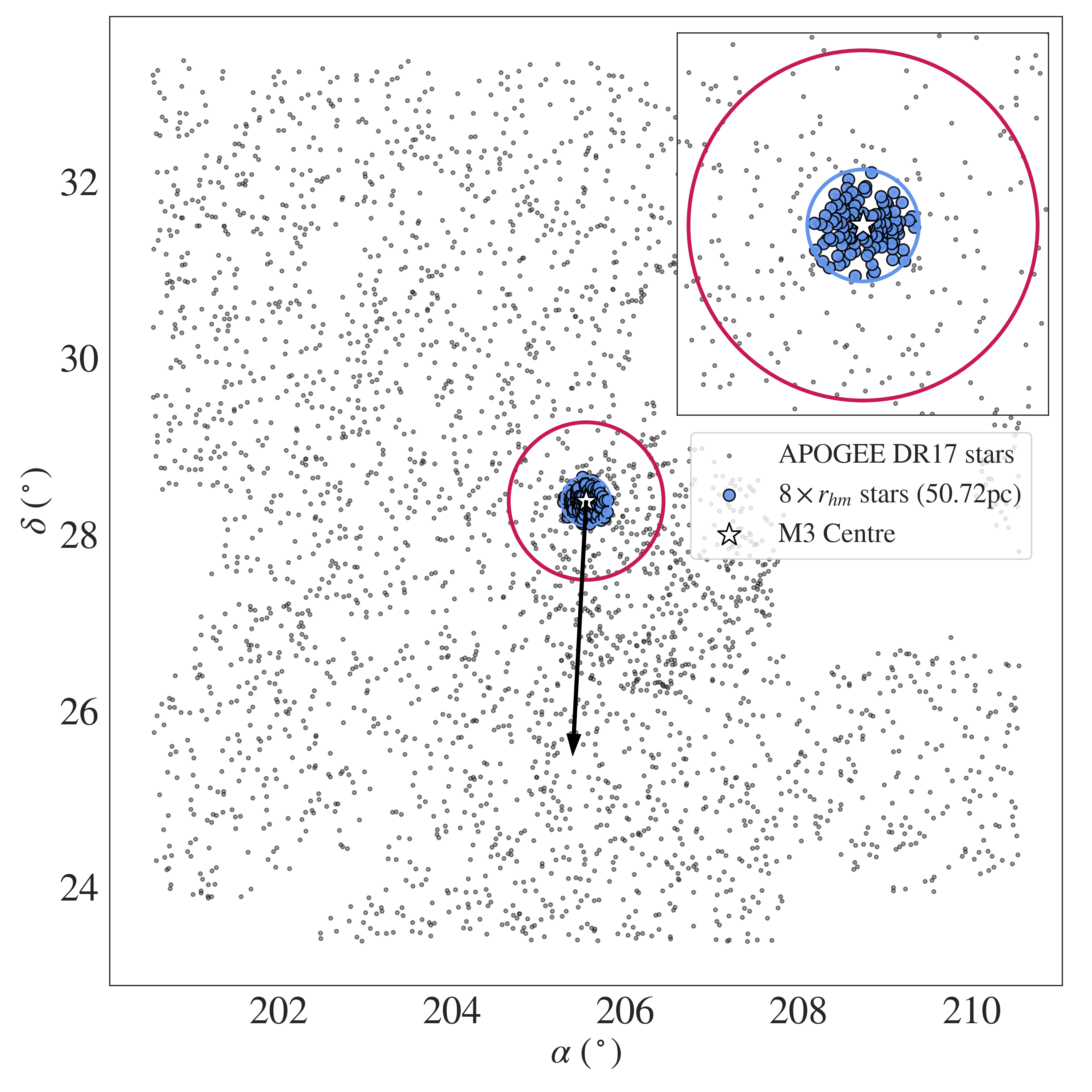}
    \caption{Distribution of 3212 APOGEE DR17 stars in a $10^{\circ} \times 10^{\circ}$ FOV around M3. The centre of M3 ($205.548^{\circ}, 28.377^{\circ}$) is marked with a white star. Stars that are within eight times the  half-mass radius (blue circle) of $8 \times r_{hm} = 50.72$pc are indicated as blue points. M3's tidal radius at apogalacticon of $r_{t} = 159.03$pc is shown as a magenta circle and cluster proper motion is indicated with a black arrow. The right ascension, declination and proper motion of M3 are all obtained from \citet{Vasiliev2019} whereas $r_{hm}$ is obtained from \citet{2018MNRAS.478.1520B} and $r_{t}$ at apogalacticon is computed using Equation 8 in \citet{2013ApJ...764..124W}.}
    \label{fig:clusterstars}
\end{figure}

Once this initial sample has been established, we define members of M3 to act as a control group for future analyses and comparisons. Specifically, we wish to identify members that are (i) spatially aligned with the reported cluster centre, (ii) chemically similar and (iii) have similar radial velocities. For members of a given GC, all three of the aforementioned parameters should be similar. While the latter two conditions are applied via machine learning clustering algorithms in Section \ref{sec:observational}, we initially select stars that are within eight times the \cite{2018MNRAS.478.1520B} M3 half-mass radius of $r_{hm} = 6.34$pc $\sim 0.036^{\circ}$. The spatial constraint of $8 \times r_{hm}$ is chosen as it yields a control group of 133 cluster stars that contain a sufficient amount of stars within the \cite{2013ApJ...764..124W} tidal radius at apogalacticon of $r_{t} = 159.03$pc. It is important to note that we use M3's tidal radius at apogalacticon to only select extra-tidal stars that are located beyond the tidal radius at \textit{all} points along M3's orbit of the Galaxy. In other words, this control group contains approximately $45\%$ of the stars within the tidal radius of M3, so contamination from field stars is expected to be minimal. Figure \ref{fig:clusterstars} highlights the spatial distribution of APOGEE DR17 stars and spatially-identified cluster members of M3 used in this study.

\begin{figure*}
    \centering
    \includegraphics[width=\textwidth]{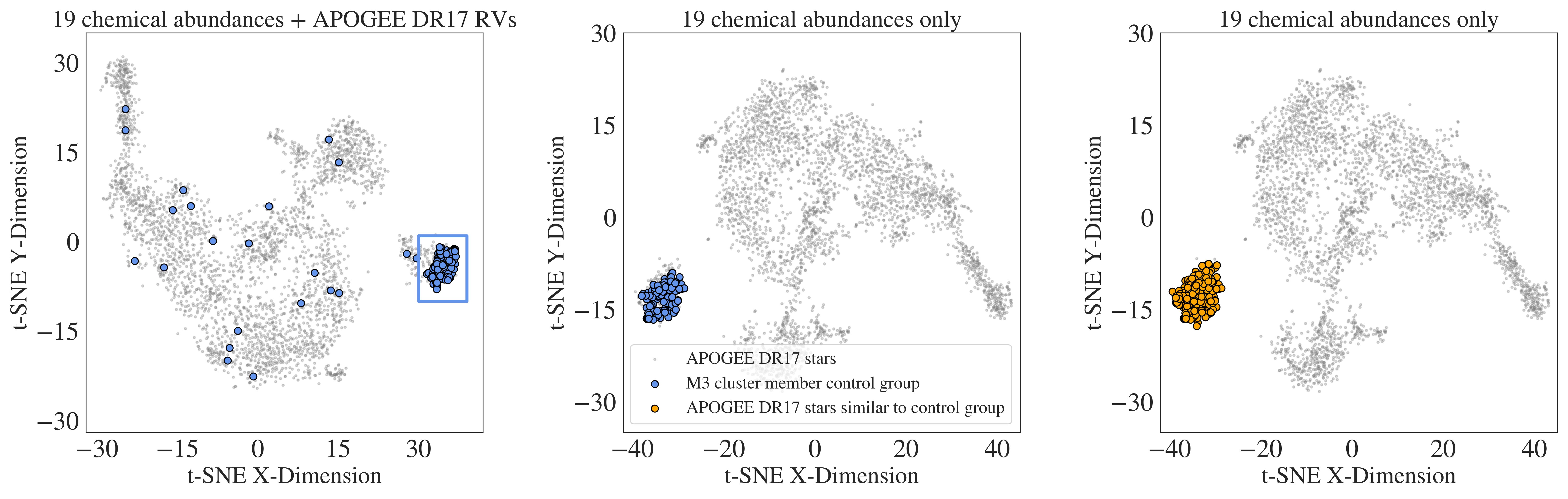}
    \caption{T-SNE projections of 3212 APOGEE DR17 stars (gray points) in a $10^{\circ} \times 10^{\circ}$ FOV around M3. With the exception of setting perplexity=100, default t-SNE input parameters are utilized. The left panel highlights the results of a t-SNE reduction of 19 \texttt{astroNN} chemical abundances and APOGEE DR17 radial velocities. A group of 133 stars within $8 \times r_{hm}$ of M3 are highlighted as blue points. To define a stronger control group, we only select cluster members within the blue box, as in addition to spatial proximity, they share similar chemical abundances and radial velocities. As extra-tidal stars may not have similar radial velocities to the suspected parent cluster, we plot the results of a new t-SNE reduction \textit{without} radial velocities (only 19 chemical abundances) in the middle panel. This panel highlights the aforementioned updated control group of 111 stars of M3 along with APOGEE DR17 stars. The right panel is the same as the middle panel, however it highlights 232 APOGEE DR17 stars that have similar chemical abundances to the control group of M3 stars (orange points).}
    \label{fig:tsne}
\end{figure*}

\section{Methodology}\label{sec:methods}

We present a new method to identify extra-tidal stars of Galactic GCs that encompasses both observations and theory. By identifying stars that are chemically similar to cluster members through high-order dimensional analysis, we produce a sample of observational extra-tidal candidates. To confirm these candidates, we use \texttt{corespray} — a new Python-based particle spray code that uses three-body dynamics to simulate the creation of extra-tidal stars. We further assess the origins of each extra-tidal candidate by computing the probability that each star belongs to a distribution of (i) simulated \texttt{Corespray} extra-tidal stars of M3 and (ii) field stars around M3. The complete methodology is outlined below and can be applied to search for extra-tidal stars of any GC in the Milky Way. 

\subsection{High-Order Dimensional Analysis}\label{sec:observational}

\subsubsection{Analysis with the t-SNE algorithm}\label{sec:tsne}

While we have spatially identified a control group of 133 cluster members of M3, we must ensure that these stars are also similar in chemical abundances and radial velocities. Doing so further mitigates the chances of unintentionally including background or foreground field stars in our control group. However, identifying similarities in 19 \texttt{astroNN} abundances plus APOGEE DR17 barycentric radial velocities requires the use of high-dimensional data analysis. We therefore adopt the methodology from \cite{2020ApJ...900..146C} and use the t-Stochastic Neighbour Embedding (t-SNE) algorithm \citep{JMLR:v9:vandermaaten08a} in the \texttt{scikit-learn} Python package \citep{scikit-learn}. T-SNE is a dimensionality-reducing clustering algorithm that finds similarities between data points in high-dimensional parameter spaces and projects them into a two-dimensional plane. Although the t-SNE plane does not correspond to a physical parameter space, it generally groups stars that share similarities in all dimensions of the input data in similar locations in the reduced two-dimensional space. 

We run the t-SNE algorithm on both the cluster members and APOGEE DR17 stars in our FOV with 19 chemical abundances and radial velocities to narrow down our cluster member control group. In turn, this step allows us to obtain a new and improved control group of cluster members that are not only spatially similar, but chemically and kinematically similar as well.  It is important to note that while \cite{1978ApJ...223..487C} and \cite{1980ApJ...237L..87P} have shown that GCs can contain multiple stellar populations, we treat cluster members as a single population with similar chemical abundances in this study. Regardless, it is straightforward to generalize this dimensionality reduction to examine other potential stellar populations.

Although t-SNE is a powerful clustering tool, it is critical to understand that output t-SNE parameter spaces are greatly impacted by (i) changing the input t-SNE parameters (ii) skewed data and (iii) re-running the algorithm (although one can set the initial condition by the random seed used for the data range). For this application, we use default t-SNE input parameters with the exception of perplexity ($p$). Perplexity is a relative weight between local and global structure in the data \citep{wattenberg2016how}. For large high-dimensional data sets, it is recommended to use higher perplexity values to yield tighter clustering, as data becomes much sparser in higher dimensions. As such, it makes sense that after running t-SNE with four different perplexity values ($\text{p}=2$, $\text{p}=5$, $\text{p}=10$ and $\text{p}=100$), we find that $\text{p}=100$ yields the tightest grouping of likely cluster members and is consequently implemented for the duration of this study. We also standardize our data to have $\mu=0$, $\sigma=1$ for all parameters to ensure that one parameter does not dominate or skew the clustering during the dimensionality reduction. Finally, due to a degree of randomness present in the t-SNE algorithm, locations of data (and thus clusters of data) in the two-dimensional output parameter space can vary from run to run. Although the data locations may change, our control group of cluster members always allows us to select APOGEE DR17 stars that are similar to M3, no matter how the data is distributed in a given t-SNE parameter space. \newpage

The left panel of Figure \ref{fig:tsne} highlights spatial members within $8 \times r_{hm}$ of M3 relative to the APOGEE DR17 stars in a t-SNE reduction of 19 \texttt{astroNN} chemical abundances and APOGEE DR17 radial velocities. One can observe that almost all stars in our initial M3 control group are clustered tightly in t-SNE space (boxed), so although there is likely some contamination by foreground and background stars in this initial sample, they fall outside of the clustered region and are thus discarded from the M3 control group in subsequent steps.

Before using t-SNE to identify APOGEE DR17 stars similar to our cluster member control group, we must note that while members of a particular GC have similar radial velocities, extra-tidal stars may not. Recall from Section \ref{sec:intro} that extra-tidal stars are created during core three-body interactions, when single stars receive velocity kicks. Depending on the configuration of the three-body system, the velocity imparted to the single star can be high. Furthermore, extra-tidal stars escape at various points along a GC's orbit, allowing for escaped stars of the same cluster to experience different phenomena (e.g. gravitational potentials, interactions with other objects, etc.). In fact, the orbital phase of the "escaper" star (i.e. the one that actually escapes the GC's gravitational pull) alone can result in highly different radial velocities. So while extra-tidal stars likely had similar radial velocities as the parent GC when they first originated, they do not necessarily have similar radial velocities at present. 

To identify extra-tidal candidates that are similar to the parent cluster, we re-run the t-SNE algorithm on the same control group of cluster members boxed in the left panel of Figure \ref{fig:tsne} and APOGEE DR17 stars \textit{without} radial velocities. The middle panel in Figure \ref{fig:tsne} highlights the control group in a t-SNE reduction containing only the 19 \texttt{astroNN} chemical abundances. From here, we select every APOGEE DR17 star that is within a tolerance level of two t-SNE units to a previously confirmed cluster member in the t-SNE two-dimensional plane. The right panel in Figure \ref{fig:tsne} depicts our final sample of 232 t-SNE selected stars with similar chemical abundances to M3. If we only include stars located beyond M3's tidal radius at apogalacticon, \textit{we observationally identify 103 extra-tidal candidates with the t-SNE algorithm}.

\begin{figure*}
    \centering
    \includegraphics[width=\textwidth]{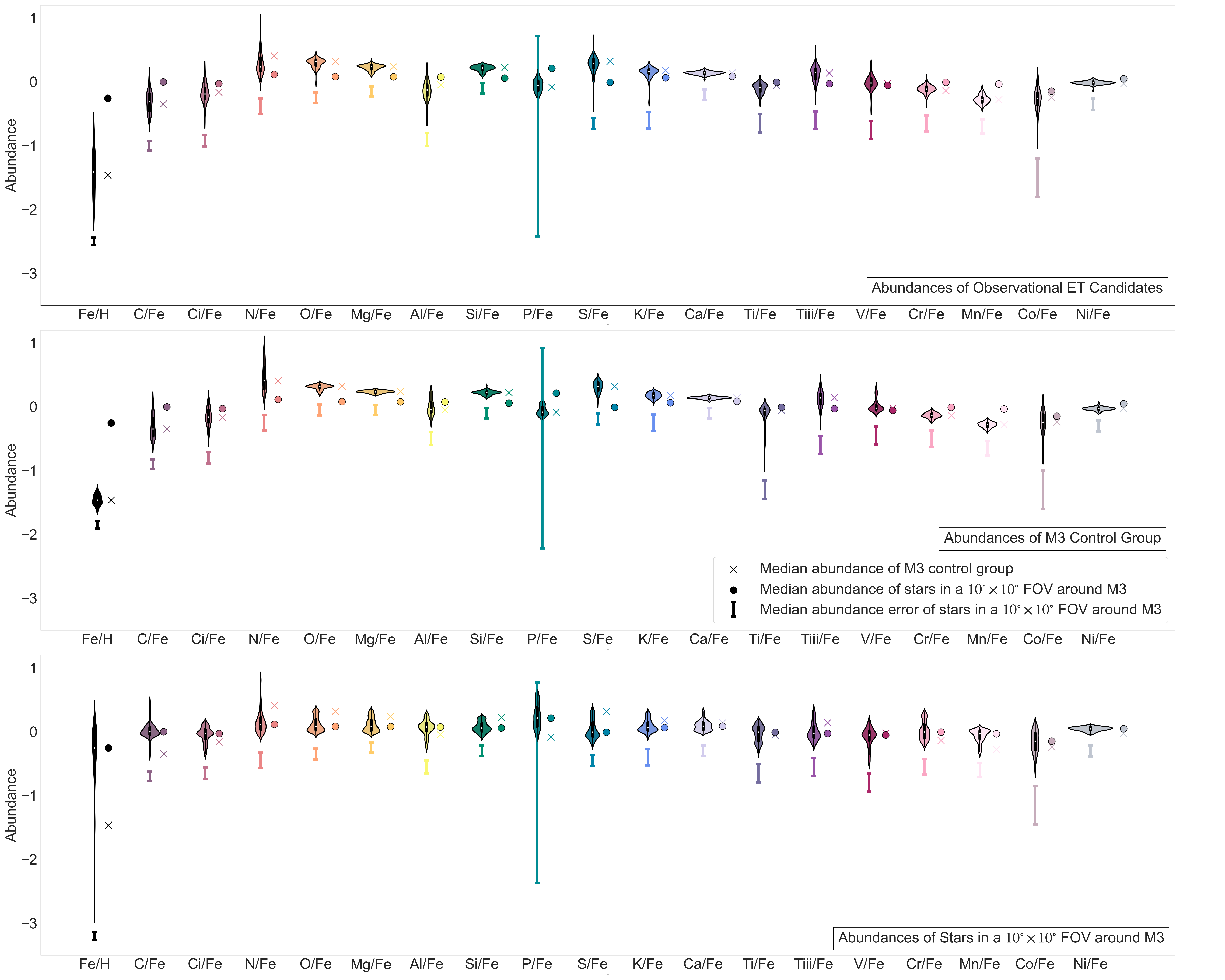}
    \caption{Abundance distributions for 103 observational extra-tidal candidates (top panel), the M3 control group (middle panel) and stars in the entire $10^{\circ} \times 10^{\circ}$ FOV around M3 (bottom panel). Outlier stars in the M3 control group and the full $10^{\circ} \times 10^{\circ}$ sample are removed by selecting only the central 95\% of data to be displayed. The violin plots represent the abundance distributions of all 19 elements in each of the aforementioned samples. The median abundances of each element in the M3 control group are marked with crosses in each panel. The median abundances and abundance errors of stars in the full $10^{\circ} \times 10^{\circ}$ FOV are indicated with circles and error bars, respectively. Each element used in the t-SNE and UMAP reduction is assigned the same colour for the violin, median abundance cross and median error bar. Comparing the violin distributions and median abundance locations in the top and middle plots, it is clear that the 103 observational extra-tidal candidates have similar abundances to those of the M3 control group. The bottom plot highlights M3's unique chemical location relative to the field stars in our $10^{\circ} \times 10^{\circ}$ sample.}
    \label{fig:violin}
\end{figure*}

\subsubsection{Analysis with the UMAP algorithm}

To further confirm our t-SNE-identified extra-tidal candidates, we perform the same high-dimensional analysis with another machine learning approach — Uniform Manifold Approximation and Projection (UMAP). Developed by \cite{2018arXiv180203426M}, UMAP is also a dimensionality-reducing clustering algorithm. However, despite its similarities to t-SNE, UMAP better preserves the global structure of the high-dimensional data, permits the addition of new data and allows embedding in arbitrary dimensions (rather than just two-dimensions). As such, it is a useful tool to identify extra-tidal candidates in large data sets like APOGEE DR17.

We follow the same process as outlined in Section \ref{sec:tsne}; first identifying a control group of cluster members based on chemical abundances and radial velocities, then selecting stars that are similar to the cluster members with just chemical abundances. The initial UMAP analysis that includes radial velocities was run for a range of input parameters, including the size of the local neighbourhood, the minimum distance between points, the number of components of the reduced dimensionality space and choice of the metric parameter. We find that a neighbourhood size of 15, minimum distance of 0.1, and the canberra metric result in M3 candidates being clearly separated from the field stars when reducing the data down to two dimensions (and as such, the selection of chemically similar stars did not require a UMAP tolerance parameter). Repeating the analysis without radial velocities allows us to \textit{observationally identify 119 extra-tidal candidates with the UMAP algorithm.} When we cross-match our t-SNE extra-tidal candidates with those identified by UMAP, we obtain a final observational sample of \textit{103 observationally-identified extra-tidal candidates}. The discrepancy between methods is most likely due to some stars not being as close to cluster members in each of the t-SNE and UMAP regimes. Regardless, almost all UMAP extra-tidal candidates are also identified with the t-SNE algorithm, providing confidence in this high-dimensionality approach to identifying extra-tidal candidates of M3. 

The 19 chemical abundance distributions for the observational extra-tidal candidates, the M3 control group and stars in the entire $10^{\circ} \times 10^{\circ}$ FOV around M3 are also presented as violin plots in Figure \ref{fig:violin}. Outlier stars in the M3 control group and the full $10^{\circ} \times 10^{\circ}$ sample are removed by selecting only the central 95\% of the data. From the violin shapes, one can observe that the abundance distributions of the extra-tidal candidates are similar to those of M3, which is consistent with those stars originating in that cluster. The median elemental abundances for both the M3 control group and stars in the full $10^{\circ} \times 10^{\circ}$ sample are also indicated in the three datasets, which again highlight both the chemical similarities between the observational extra-tidal candidates and M3 and the unique chemical composition of M3 relative to the surrounding field stars. Finally, the median abundance error for each element in the $10^{\circ} \times 10^{\circ}$ sample is presented in all three plots to show the typical error sizes for each element used in this analysis. These errors are taken directly from the \texttt{astroNN} abundance catalogue \citep{2019MNRAS.483.3255L, 2019MNRAS.489.2079L}.

\subsection{\texttt{Corespray} Particle Spray Code}\label{sec:theoretical}

While t-SNE and UMAP are useful tools in identifying stars beyond a cluster's tidal radius that are chemically similar to GC members, they neglect to provide information on whether a given star could have escaped a GC. Furthermore, chemical similarities alone do not necessarily indicate that stars were born in the same GC \citep{2018ApJ...853..198N, 2019ApJ...883..177N, 2020MNRAS.494.2268W, 2021A&A...654A.151C}. Thus, information on where extra-tidal stars of a given GC could end up in a variety of parameter spaces is an essential tool that is necessary for confirming the extra-tidal nature of observationally-identified extra-tidal candidates.

To further constrain our extra-tidal candidates and probe three-body dynamics in the cores of GCs, we present \texttt{corespray}\footnote{For a complete description of \texttt{corespray}'s installation instructions and capabilities, please visit \url{https://github.com/webbjj/corespray}.} — a Python-based particle spray code that uses three-body dynamics to simulate statistical samples of extra-tidal stars for any Galactic GC. \texttt{Corespray} will simulate $N$ extra-tidal stars, only requiring the user to input a set of conditions unique to their GC of interest (e.g. mass, core density, binary fraction, central velocity dispersion, escape velocity, etc.). These GC parameters can almost always be found in \cite{2018MNRAS.478.1520B}.

To simulate extra-tidal stars, \texttt{corespray} initially defines a three-dimensional position and velocity within the cluster for single stars and binaries of three-body systems at a random time along the orbit of the GC around the Galaxy. It is important to note that \texttt{corespray} does not simulate triple systems (i.e. bound three-body systems composed of three single stars and no binary), as the presence of these systems in GCs is believed to be rare \citep{2011MNRAS.416.1410L}. All orbits in \texttt{corespray} are defined and integrated using \texttt{galpy} \footnote{For a complete description of \texttt{galpy}, please visit \url{http://github.com/jobovy/galpy}.} — a Python-package for galactic dynamics \citep{2015ApJS..216...29B}. For each three-body simulation, masses of the single star ($m_{s}$) and recoil binary ($m_{a}$, $m_{b}$) are sampled from a power-law distribution with a slope of $\alpha=-1.35$ \citep{1955ApJ...121..161S} and a mass range between $ 0.08 M_\odot < m < 1.4 M_\odot$. The system masses determine the probability that a single star escapes the three body system ($P_{s}$) and is computed in Equation \ref{eq:prob} \citep{3body}.

\begin{equation}
P_{s} = \dfrac{m_{s}^{-3}}{m_{s}^{-3} + m_{a}^{-3} + m_{b}^{-3}}
\label{eq:prob}
\end{equation}

By randomly sampling the probability function, \texttt{corespray} determines if the star escapes the system and if so, computes the system's total energy. The total energy of the three-body system ($E_{0}$) is computed by summing the kinetic and gravitational potential energies of both the single kicked star ($E_{s}$) and the recoil binary ($E_{B}$). To compute $E_{B}$, we first recognize that binaries with circular orbits have constant velocity, so $\dot{r} = 0$. Thus, $E_{B}$ is totally dependent on the gravitational potential energy of the binary which is sampled between twice the hard-soft boundary and twice the contact boundary between two solar mass stars. To compute both the kinetic and gravitational potential energy of $E_{s}$, we require the position vector between (i) the single star and the centre of mass of the binary ($\mathbf{r}_{s}$) and (ii) the reduced mass of the motion relative to the single star ($m = \frac{m_{B} m_{s}}{M}$). The total mass of the three-body system ($M$) is represented as $M = m_{s} + m_{B}$. Thus, we sum the components of $E_{s}$ and $E_{B}$ to compute $E_{0}$ in Equation \ref{eq:en}, where $\dot{\mathbf{r}}_{s}$ is with respect to the reference frame of the binary \citep{3body}.

\begin{equation}
    E_{0} = \frac{1}{2} m \dot{\mathbf{r}}_{s}^{2} - G \frac{m_{s} m_{B}}{r_{s}} + E_{B}
    \label{eq:en}
\end{equation}

 \noindent With $E_{0}$, \texttt{corespray} computes the escape velocity distribution of the star $f(v_{s})$ via Equation \ref{eq:vdist} \citep{3body}.

\begin{equation}
f(v_{s}) dv_{s} = \dfrac{(3.5|E_{0}|^{7/2} m_{s} M/m_{B}) v_{s} d v_{s}}{(|E_{0}| + \frac{1}{2} (m_{s} M/m_{B}) v_{s}^{2})^{9/2}}
\label{eq:vdist}
\end{equation}

 \noindent By computing the maximum of Equation \ref{eq:vdist} via $\frac{df}{d v_{s}}=0$, one can solve for the peak escape velocity $v_{s, peak}$ through Equation \ref{eq:vpeak} \citep{3body}.

\begin{equation}
v_{s, peak} = \frac{1}{2} \sqrt{\frac{(M - m_{s})}{m_{s} M}} \sqrt{|E_{0}|}
\label{eq:vpeak}
\end{equation}

\begin{figure}
    \centering
    \includegraphics[width=0.85\columnwidth, height=18cm]{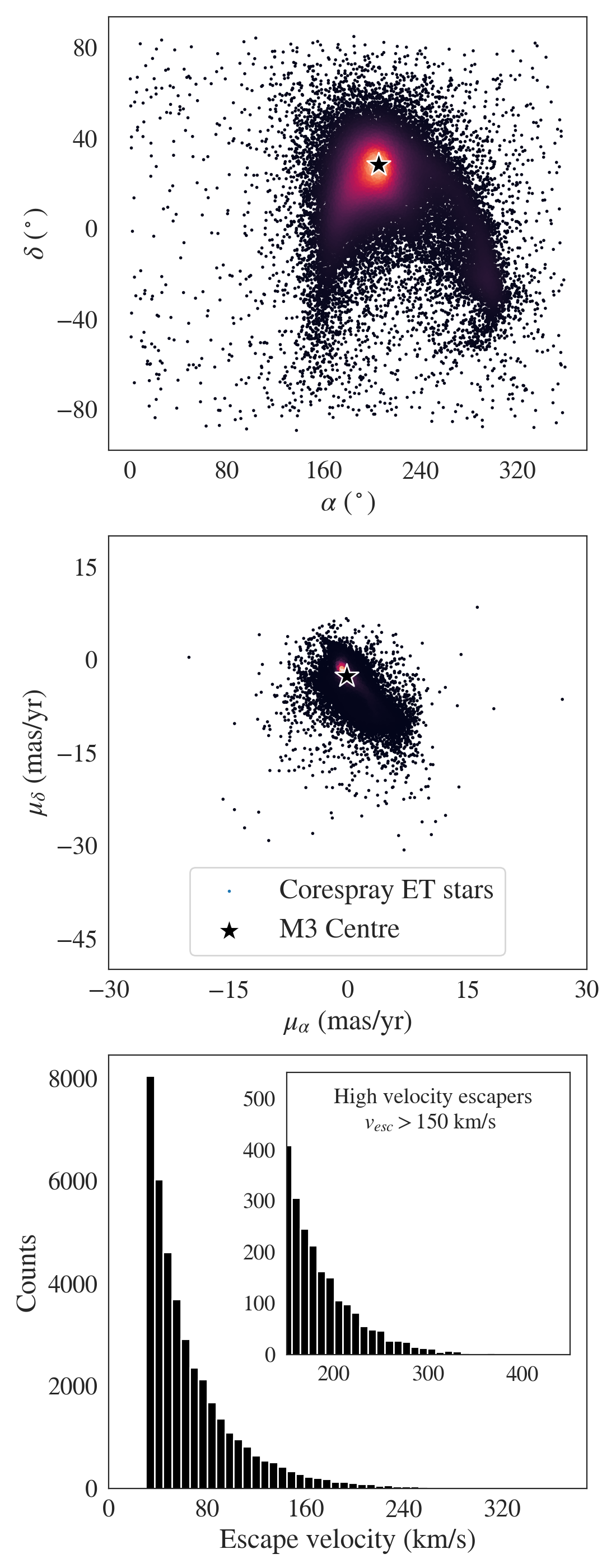} 
\caption{Spatial (top panel), proper motion (middle panel) and escape velocity (bottom panel) distributions of 40,000 extra-tidal stars of M3 simulated with \texttt{corespray}. Each extra-tidal star escaped the core of M3 at a random time during one azimuthal orbital period ($\sim 387$ Myr) of the GC around the Galaxy. Spatial locations and proper motions are shown as scatter density representations with a kernel-density estimate using Gaussian kernels from \texttt{scipy.stats.gaussian\_kde}. The location and proper motion of M3 is indicated with a black star in both plots respectively. Escape velocities for all simulated stars are depicted in a histogram with a sub-panel that highlights the high velocity escapers (i.e. $v_{\text{esc}} > 150$ km/s). One can observe that while the locations and proper motions of extra-tidal stars of M3 are most concentrated near the cluster itself, extra-tidal stars can leave the cluster with a wide variety of escape velocities and proper motions.}
    \label{fig:cspray}
\end{figure}

Through randomly sampling the escape velocity distribution function in Equation \ref{eq:vdist} between $ 0 < v_{s, test} < 5 \times v_{s, peak}$ (Equation \ref{eq:vpeak}), we determine that the kicked star is extra-tidal only if the sampled test velocity is larger than the \cite{2018MNRAS.478.1520B} escape velocity of the GC. \texttt{Corespray} will continue to sample three-body interactions until $N$ escaper stars are produced. Once $N$ extra-tidal stars are simulated, \texttt{corespray} projects the escape velocities onto the initial three-dimensional velocity vectors to determine the directions of motion of the escaper stars. Escaper positions and velocities are updated and new orbits are defined for each star. It is important to note that \texttt{corespray} assumes that each cluster is isotropic and thus has no rotation. Combined with the fact that \texttt{corespray} ejects stars at various times along the GC's orbit throughout the Galaxy, this assumption ensures that the stars kick velocity vector is oriented in a random direction.

Although integration time is a free parameter in \texttt{Corespray}, simulating recent escapers of GCs produces  extra-tidal star distributions that are therefore minimally affected by Galactic substructure or differences between the true Galactic tidal field and an assumed model. Thus, we simulate extra-tidal stars that have escaped over the duration of one azimuthal period. After escaping the cluster at some random time during the GC's period, the simulated stars are integrated from their escape time to the present day in a combined Milky Way and King potential. While the potentials are also free parameters in \texttt{corespray}, we specifically use \texttt{MWPotential2014} and \texttt{KingPotential} from \texttt{galpy} \citep{2015ApJS..216...29B} to encompass influences from both the Galaxy and the GC, respectively. Along with three-dimensional positions and velocities, escape times and velocities for all the simulated stars are computed in \texttt{corespray}. With these quantities, a variety of orbital parameters (i.e. right ascensions, declinations, proper motions, radial velocities and distances) can be computed with \texttt{galpy} \citep{2015ApJS..216...29B}. Ultimately, \texttt{corespray} allows the user to define a variety of parameter spaces to obtain a statistical representation of single extra-tidal stars and binary extra-tidal systems for a given GC. Figure \ref{fig:cspray} highlights three example parameter spaces generated from a \texttt{corespray} simulation of $40,000$ extra-tidal stars of M3.

\subsection{Extra-Tidal Candidate Probabilities}\label{sec:confirmation}

In Section \ref{sec:observational}, we use high-order dimensional analysis to identify 103 extra-tidal candidates that are both spatially and chemically similar to a control group of M3 stars. While t-SNE and UMAP are successful at identifying candidate stars isolated in the field, these algorithms just confirm that the stars share similar chemistry across 19 different elements. Thus, in Section \ref{sec:theoretical} we develop \texttt{corespray} to further constrain the \textit{origin} of our observationally-identified extra-tidal candidates. However, quantifying the probability that each extra-tidal candidate is an extra-tidal star of M3 requires us to infer two distributions: one distribution of extra-tidal stars of M3 (\texttt{corespray}) and another distribution of field stars surrounding M3.

To investigate our extra-tidal candidates' origins, we first use \texttt{corespray} to simulate 40,000 extra-tidal stars of M3 over the duration of one azimuthal orbital period of $P_{\text{orb}}$ = $387.82$ Myr (Figure \ref{fig:cspray}). The input parameters used in this simulation are summarized in Table \ref{tab:inputparams}. Since we restrict our observational search to a $10^{\circ} \times 10^{\circ}$ FOV, we also constrain our sample of \texttt{corespray} extra-tidal stars to be confined within this spatial range (3878 stars). In a two-dimensional distribution ($\alpha$ vs.  $\delta$), we find that 484 of these stars lie inside M3's tidal radius. While some of these stars could have low kick velocities and are still in the process of escaping the cluster, most have high radial velocities and only appear to reside inside the cluster due to projection effects. Regardless we  remove these stars from our \texttt{corespray} distribution for consistency. Altogether, this yields a sample of 3394 simulated extra-tidal stars that can be used to probe cluster associations for each extra-tidal candidate. It is important to note that \texttt{corespray} extra-tidal stars located within a $10^{\circ} \times 10^{\circ}$ FOV around M3 are ones that either received low velocity kicks or escaped the cluster recently, as they are still in proximity to the GC. Stars within this FOV represent a small sample of the full 40,000 star simulation, indicating that the majority of core interactions that could have occurred over the past orbital period result in extra-tidal stars being kicked beyond a $10^{\circ} \times 10^{\circ}$ FOV of M3. As mentioned in Section \ref{sec:data}, the kinematics of these faraway stars would likely be more affected by the Galaxy's tidal field or substructure, making it more difficult to associate these stars with M3.

\begin{table}
  \centering
  \begin{tabular}{|c c|}
  \hline
     \multicolumn{2}{|c|}{M3 \texttt{corespray} input parameters} \\
    \hline
     $\mu_{0}$ & 0.0 km/s \\
     $\sigma_{0}$ & 7.60 km/s \\
     $v_{\text{esc}, 0}$ & 30.0 km/s \\
     $\log{\rho_{0}}$ & 3.67 $M_{\odot}$ / $\text{pc}^{3}$ \\
     $r_{t}$ & 159.03 pc \\
     $\text{M}_{\text{GC}}$ & $4.06 \times 10^{5} M_{\odot}$  \\
     $m_{\text{min}}$ & 0.10 $M_{\odot}$ \\
     $m_{\text{max}}$ & 1.40 $M_{\odot}$ \\
     $\alpha$ & -1.35 \\
     $W_{0}$ & 8.61 \\
     \hline
\end{tabular}
\caption{Summary of the input parameters used in the \texttt{corespray} M3 extra-tidal star simulations. M3's average 1D velocity dispersion ($\sigma_{0}$), core escape velocity ($v_{\text{esc, 0}}$), logarithm of the core density ($\log{\rho_{0}}$) and cluster mass ($\text{M}_{\text{GC}}$) are all obtained from the \citet{2018MNRAS.478.1520B} catalogue. We assume an average 1D core velocity of $\mu_{0} = 0$ km/s and compute the tidal radius at apogalacticon ($r_{t}$) from the \citet{2013ApJ...764..124W} tidal radius at perigalacticon ($r_{t} = 127.28$pc). Minimum stellar mass ($m_{\text{min}}$) and maximum stellar mass ($m_{\text{max}}$) of stars undergoing the three-body encounters in the core are chosen to include the entire mass spectrum of stars and white dwarfs. The slope of the stellar mass function in the core ($\alpha$) comes from \citet{1955ApJ...121..161S}. Finally, the King central potential parameter derives from $W_{0} = \log{r_{t} / c}$ where c is the concentration obtained from \citet{Harris2010}.}
  \label{tab:inputparams}
\end{table}

From our \texttt{corespray} simulation, we have a sample of extra-tidal stars of M3. In contrast, we must produce a distribution of stars that are clearly not extra-tidal stars of M3 for comparison. Thus, we define a "field star" distribution by removing all sources from our original $10^{\circ} \times 10^{\circ}$ FOV sample that are either (i) members of the M3 control group defined in Section \ref{sec:observational} or (ii) any of the 103 extra-tidal candidates. Our field star distribution contains 2998 stars. With these two samples, we can now compute the probability that each extra-tidal candidate belongs to the \texttt{corespray} extra-tidal star distribution or field star distribution. To do this calculation, we incorporate proper motions and radial velocities into both a multivariate Gaussian distribution model and an extreme deconvolution (XD). 

To compute the probabilities of extra-tidal candidates belonging to the \texttt{corespray} distribution, we model the \texttt{corespray} data with a multivariate Gaussian distribution in \texttt{scikit-learn} \citep{scikit-learn}. Using Gaia EDR3 proper motions and APOGEE DR17 radial velocities, a mean and covariance matrix can be generated and used to construct the multivariate Gaussian. To ensure accuracy in the covariance, the covariance matrix must be computed for each extra-tidal candidate by summing the covariance matrix from \texttt{corespray} and the covariance matrix from the extra-tidal candidate distribution. From this model, the probability distribution function (PDF) can be computed and individual extra-tidal candidate \texttt{corespray} probabilities, $\mathbb{P}(\text{c})$, can be extracted.

 \begin{figure*}
    \centering
    \includegraphics[width=\textwidth]{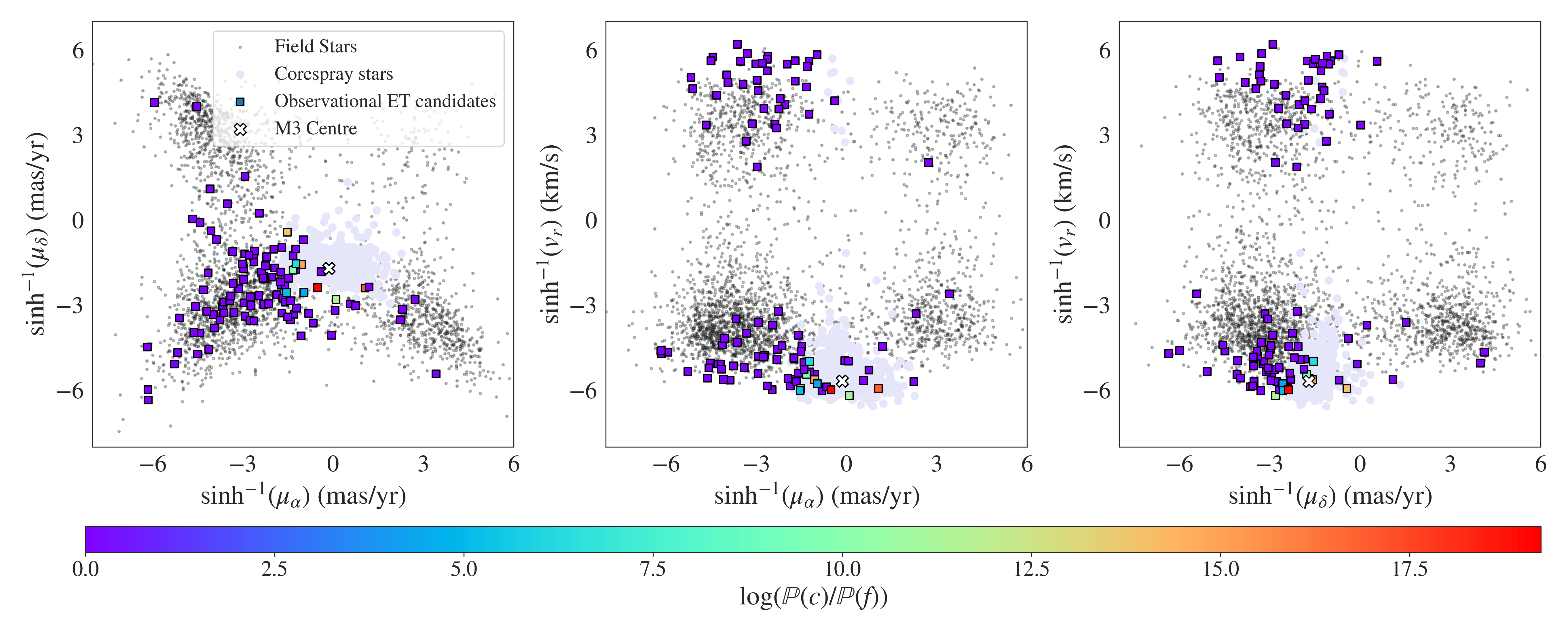}
    \caption{Proper motion and radial velocity spaces for all 103 observationally identified extra-tidal candidates (coloured squares) relative to 2998 field stars (black points) and 3394 \texttt{corespray} stars (lavender circles) in a $10^{\circ} \times 10^{\circ}$ FOV around M3. Extra-tidal candidates are numbered and coloured by their $\log{(\mathbb{P}(\text{c}) / \mathbb{P}(\text{f}))}$ value, where stars with $\log{(\mathbb{P}(\text{c}) / \mathbb{P}(\text{f}))} < 0$ are treated equally and assigned the same colour (dark purple). This is done to highlight stars with positive $\log{(\mathbb{P}(\text{c}) / \mathbb{P}(\text{f}))}$ values, as they have higher probabilities of belonging to the \texttt{corespray} distribution than the field. Each representation depicts the $\sinh^{-1}$ of the data, as it best separates low and high values to allow for optimal data visualization. M3's proper motion of $(\mu_{\alpha}, \mu_{\delta}) = (-0.142, -2.647) \ \text{mas/yr}$ and radial velocity of $v_{r} = -147.28 \ \text{km/s}$ are obtained from \citet{Vasiliev2019} and used to indicate the centre of M3 (white cross) in each panel.}
    \label{fig:logodds}
\end{figure*}

Contrary to the simulated \texttt{corespray} distribution, the field star distribution is observed and thus subject to measurement error. Unfortunately, the previous multivariate Gaussian modelling is unable to directly incorporate error when defining the distribution and thus must be modified for noisy data sets. However, \cite{2011AnApS...5.1657B} outline an approach that can analyze noisy, heterogeneous and incomplete data; specifically, an XD. 

\begin{figure*}
    \centering
    \includegraphics[width=\textwidth]{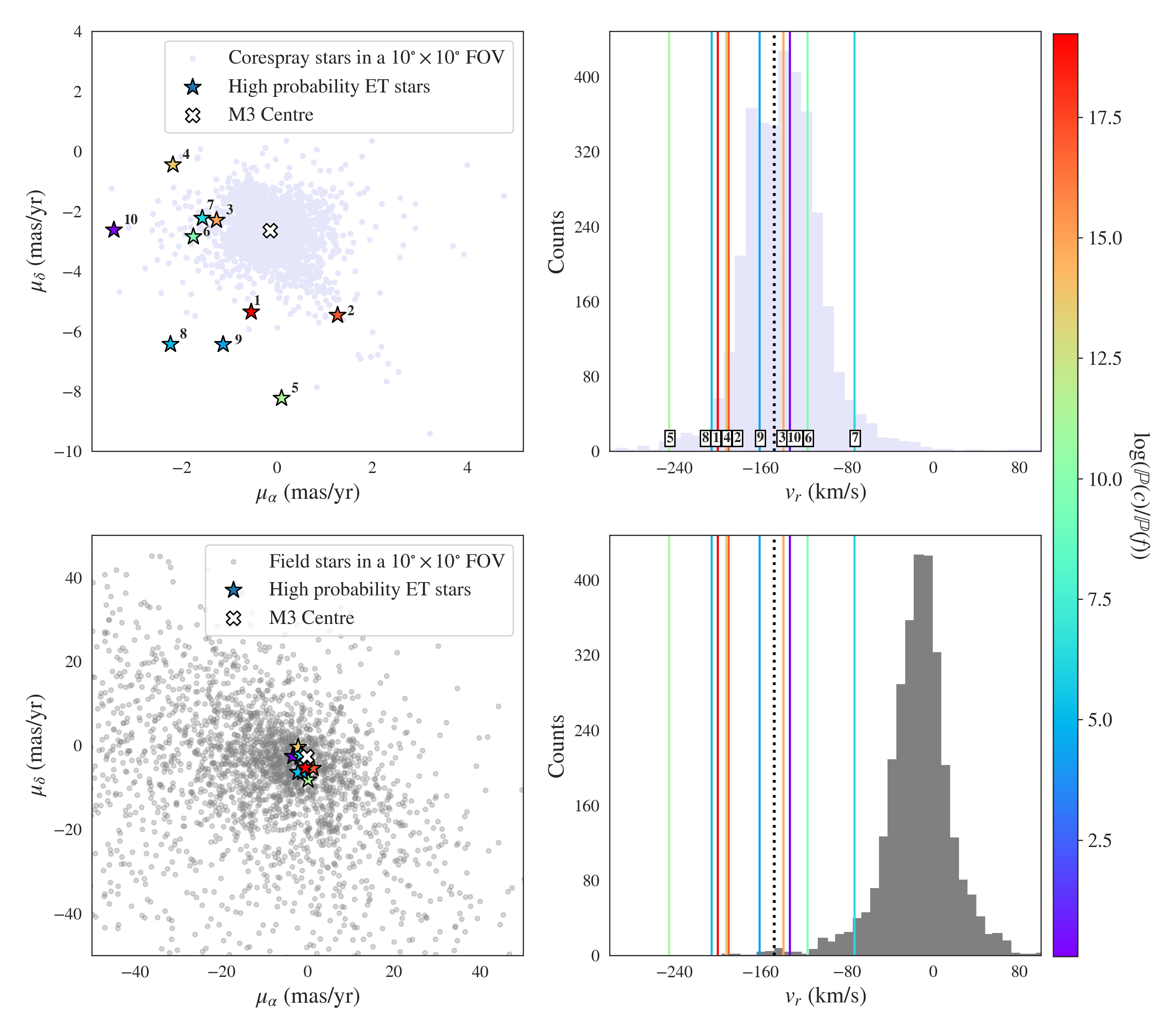}
    \caption{Proper motion and radial velocity distributions of 10 high-probability extra-tidal stars relative to a sample of 3394 \texttt{corespray} extra-tidal stars in a  $10^{\circ} \times 10^{\circ}$ FOV around M3 (lavender points and distribution) and a sample of 2998 field stars in a $10^{\circ} \times 10^{\circ}$ FOV around M3 (gray points and distribution). Left panels: proper motions of the 10 new extra-tidal stars are indicated as stars coloured by their $\log{(\mathbb{P}(\text{c}) / \mathbb{P}(\text{f}))}$ values. M3's \citet{Vasiliev2019} mean proper motion of $(\mu_{\alpha}, \mu_{\delta}) = (-0.142, -2.647) \ \text{mas/yr}$ is indicated with a white cross. The proper motion field star distribution is zoomed in for clarity. Right panels: radial velocities of the 10 new extra-tidal stars are indicated as coloured lines. The colouring for each line again corresponds to the $\log{(\mathbb{P}(\text{c}) / \mathbb{P}(\text{f}))}$ of each high-probability extra-tidal star. The \citet{Vasiliev2019} mean M3 radial velocity of $v_{r} = -147.28 \ \text{km/s}$ is indicated as a dotted black line.}
    \label{fig:final}
\end{figure*}

An XD operates similarly to a Gaussian mixture model, where Bayesian estimation and Gaussian modelling output a corrected distribution \citep{astroML}. To estimate a distribution for the field stars around M3, we utilize \texttt{astroML} — a Python module for machine learning and data mining \citep{astroML}. Specifically, we use the \texttt{XDGMM} function with one Gaussian component (\texttt{n\_{components}=1}) and a default number of iterations (\texttt{max\_{iter}=100}) to perform the XD. By inputting the field star proper motions, radial velocities and associated errors into \texttt{XDGM}, we obtain a corrected field star distribution. 

The probability of each candidate star belonging to the field, $\mathbb{P}(\text{f})$, is computed using the distribution obtained from the XD in a similar manner as before. Ultimately, once $\mathbb{P}(\text{c})$ and $\mathbb{P}(\text{f})$ are computed, the odds that an extra-tidal candidate is more similar to the \texttt{corespray} sample than the field star sample can be quantified by computing the logarithm of the odds ratio: $\log{(\mathbb{P}(\text{c}) / \mathbb{P}(\text{f}))}$. The odds ratio — also referred to as the Bayes factor \citep{kassraftery} — represents the ratio of the likelihoods between two distributions and is useful to quantitatively assign a strength of association of the data to the distributions \citep{jeffreys_1935}. Ultimately, if an extra-tidal candidate has an odds ratio of $\log{(\mathbb{P}(\text{c}) / \mathbb{P}(\text{f}))} > 0$, it would indicate that the star has a higher probability of being associated with the \texttt{corespray} distribution than the field and should thus be deemed a high-probability extra-tidal star of M3. 

\section{Results}\label{sec:newstars}

For each of the 103 extra-tidal candidates we identify in Section \ref{sec:observational}, we use the methodology presented in \ref{sec:confirmation} to compute $\log{(\mathbb{P}(\text{c}) / \mathbb{P}(\text{f}))}$. To compute $\mathbb{P}(\text{c})$ via a multivariate Gaussian model, we use a mean and covariance matrix generated from the \texttt{corespray} sample. From the simulated proper motions and radial velocities, we obtain a \texttt{corespray} mean of  $\mu_{\text{c}}=( -0.153 \ \text{mas/yr}, -2.562\ \text{mas/yr}, -138.928\ \text{km/s})$. The \texttt{corespray} covariance matrix is computed by summing the covariance from the \texttt{corespray} sample (Equation \ref{eq:cov}) and the covariance of each extra-tidal candidate (Equation \ref{eq:cov2}). The extra-tidal candidate covariance values are computed from the Gaia proper motion errors and APOGEE radial velocity errors. After modelling a multivariate Gaussian distribution with these parameters, we extract $\log{\mathbb{P}(\text{c})}$ for each extra-tidal candidate.

\begin{equation}
\Sigma_{\text{c}} = 
    \begin{bmatrix}
3.636 \times 10^{-1} & -1.169 \times 10^{-1} & -3.683 \\
-1.169 \times 10^{-1} &  6.793\times 10^{-1} &  1.040 \times 10^{1} \\
-3.683 &  1.040 \times 10^{1} &  1.375\times 10^{3}  \\
\end{bmatrix}
\label{eq:cov}
\end{equation}

\begin{equation}
\Sigma_{\text{ET}} = 
    \begin{bmatrix}
\sigma_{\mu_{\alpha, i}}^{2} & 0 & 0\\
0 &  \sigma_{\mu_{\delta, i}}^{2} &  0 \\
0 &  0 &  \sigma_{v_{R, i}}^{2}  \\
\end{bmatrix}
\label{eq:cov2}
\end{equation}

To compute $\mathbb{P}(\text{f})$, we use a mean and covariance generated from the XD. As outlined in Section \ref{sec:confirmation}, we run the XD by inputting field star proper motions, radial velocities and associated errors into \texttt{XDGMM}. From this analysis, we obtain a field star mean of $\mu_{\text{f}}=(-19.804 \ \text{mas/yr}, -11.385\ \text{mas/yr}, -13.568\ \text{km/s})$ and a field star covariance outlined in Equation \ref{eq:cov3}.

\begin{equation}
\Sigma_{\text{f}} = 
    \begin{bmatrix}
3.225 \times 10^{3} & -1.721 \times 10^{1} & 1.386 \times 10^{2}\\
-1.720 \times 10^{1} &  1.672 \times 10^{3} &  -2.064 \times 10^{1} \\
1.386\times 10^{2} & -2.064 \times 10^{1} &  9.848\times 10^{2}  \\
\end{bmatrix}
\label{eq:cov3}
\end{equation}

With $\mu_{f}$ and $\Sigma_{f}$, we replicate our previous probability computation, but this time for $\mathbb{P}(\text{f})$. Like before, the final covariance matrix for the field star distribution is the sum of the covariance from the field star sample (Equation \ref{eq:cov3}) and the covariance of each extra-tidal candidate (Equation \ref{eq:cov2}). With these parameters, we construct a new multivariate Gaussian distribution, ultimately allowing us to extract $\log{\mathbb{P}(\text{f})}$ for each extra-tidal candidate. Proper motion and radial velocity distributions with $\log{(\mathbb{P}(\text{c}) / \mathbb{P}(\text{f}))}$ values are presented for each of the 103 extra-tidal candidates in Figure \ref{fig:logodds}. Although $\log{(\mathbb{P}(\text{c}) / \mathbb{P}(\text{f}))}$ ranges from $-178695.482$ to $19.233$, extra-tidal candidates with $\log{(\mathbb{P}(\text{c}) / \mathbb{P}(\text{f}))} < 0$ are treated equally and coloured the same, as these stars have higher probabilities of being associated with the field than with \texttt{corespray} extra-tidal stars. The kinematic parameter spaces are plotted by computing the $\sinh^{-1}$ of the data, as it best separates low and high values to allow for optimal data visualization.

As described in Section \ref{sec:confirmation}, our criteria for selecting high-probability extra-tidal candidates of M3 are those with $\log{(\mathbb{P}(\text{c}) / \mathbb{P}(\text{f}))} > 0$. This simply means that based on its kinematics, an extra-tidal candidate is more likely to belong to the \texttt{corespray} distribution of extra-tidal stars of M3 than a sample of field stars. Upon computation of $\log{(\mathbb{P}(\text{c}) / \mathbb{P}(\text{f}))}$ for each of our 103 chemically similar extra-tidal candidates, we find that 10 stars have $\log{(\mathbb{P}(\text{c}) / \mathbb{P}(\text{f}))} > 0$ and are thus deemed \textit{high-probability} extra-tidal stars. Of these 10 high-probability extra-tidal stars, $\log{(\mathbb{P}(\text{c}) / \mathbb{P}(\text{f}))}$ ranges from $0.081$ to $19.233$. Proper motion and radial velocity distributions for all 10 high-probability stars of M3 are presented in Figure \ref{fig:final}. The $\log{(\mathbb{P}(\text{c}) / \mathbb{P}(\text{f}))}$ values for these stars are presented in Table \ref{tab:prob}, where the stars are numbered according to their $\log{(\mathbb{P}(\text{c}) / \mathbb{P}(\text{f}))}$.  Complete spatial, kinematic and $\log{(\mathbb{P}(\text{c}) / \mathbb{P}(\text{f}))}$ information for the 103 extra-tidal candidates (including the high-probability stars) is presented in Appendix \ref{tab:allstars}.

It is interesting to note that several of the extra-tidal stars are located in the outskirts of the \texttt{corespray} proper motion and radial velocity distributions in Figure \ref{fig:final}. The high relative proper motions and radial velocities of these stars are consistent with high-velocity ejections from three-body interactions in M3. As seen in Figure \ref{fig:cspray}, three-body interactions producing high velocity ejections are rare, but not completely uncommon. Given the number of high-velocity extra-tidal stars relative to the simulated \texttt{corespray} stars, it may be the case that the core of M3 has a higher density, a flatter stellar mass function, or a sub-population of black holes such that three-body interactions are primarily between higher-mass stars and remnants than considered here. Alternatively, perhaps the central escape velocity reported in \cite{2018MNRAS.478.1520B} is too low, causing low velocity stars to remain bound to the cluster in the \texttt{corespray} simulation. As such, fewer low velocity stars would populate the \texttt{corespray} distribution, resulting in low velocity observational extra-tidal candidates having kinematics inconsistent with the \texttt{corespray} extra-tidal stars. Moreover, it is possible that some extra-tidal stars have magnitudes fainter than the APOGEE magnitude limit. Consequently, we might only be observing a bright sub-sample of all the extra-tidal stars produced from M3. Regardless, these findings further indicate that imposing proper motion or radial velocity constraints when initially searching for extra-tidal stars could result in extra-tidal candidates from high-velocity ejections being missed.

\section{Discussion}\label{sec:discussion}

\subsection{Extra-Tidal Star Probabilities}

\begin{table}
\centering
\begin{tabular}{|c|c|c|c|c|}
\hline \# & Extra-tidal Star ID & $\log{(\mathbb{P}(\text{c}) / \mathbb{P}(\text{f}))}$ & Evidence Strength \\
\hline
1 & 2M13500350+2431542 & 19.233 & Decisive \\
2 & 2M13240682+3020316 & 17.002 & Decisive \\
3 & 2M13553890+3241208 & 14.809 & Decisive \\
4 & 2M13382215+3233031 & 13.763 & Decisive \\
5 & 2M13413296+3255410 & 11.305 & Decisive \\
6 & 2M13234701+3111279 & 9.749 & Decisive \\
7 & 2M13271850+2841521 & 6.358 & Decisive \\
8 & 2M13353852+2939287 & 5.118 & Decisive \\
9 & 2M13251237+3018535 & 4.307 & Decisive \\
10 & 2M13524016+2601592 & 0.081 & NWBM \\
\hline
\end{tabular}
\caption{The $\log{(\mathbb{P}(\text{c}) / \mathbb{P}(\text{f}))}$ values for the 10 new high-probability extra-tidal stars of M3. Stars (and their corresponding APOGEE IDs) are organized in descending order of $\log{(\mathbb{P}(\text{c}) / \mathbb{P}(\text{f}))}$. The Kass \& Raftery strength of evidence metric to quantify each star's association to the \texttt{corespray} distribution is also presented \citep{kassraftery}. Values of $0<\log{(\mathbb{P}(\text{c}) / \mathbb{P}(\text{f}))}<0.5$ are not worth more than a bare mention (NWBM), $0.5<\log{(\mathbb{P}(\text{c}) / \mathbb{P}(\text{f}))}<1$ are significant, $1<\log{(\mathbb{P}(\text{c}) / \mathbb{P}(\text{f}))}<2$ are strong and $\log{(\mathbb{P}(\text{c}) / \mathbb{P}(\text{f}))}>2$ are decisive of being associated with the \texttt{corespray} extra-tidal stars.}
\label{tab:prob}
\end{table}

After spatial, chemical and kinematic analyses, we have identified 10 high-probability extra-tidal stars of the Galactic GC M3. Our metric for determining the highest-probability extra-tidal stars is simple: we only select stars that have $\log{(\mathbb{P}(\text{c}) / \mathbb{P}(\text{f}))} > 0$. However, \cite{kassraftery} famously provide metrics to interpret the logarithm of the odds ratio values and assign a strength of evidence that data is associated with a given distribution. Based on our computed $\log{(\mathbb{P}(\text{c}) / \mathbb{P}(\text{f}))}$ and the \cite{kassraftery} interpretation, we quantify the strength of evidence for each extra-tidal star in Table \ref{tab:prob}. From this, we observe that 9 stars are decisive and 1 star is not worth more than a bare mention with being associated with the \texttt{corespray} extra-tidal star distribution. It is important to mention that the above interpretation is equivocal and is meant to act more as a general guideline than a definite statement. Regardless, the positive $\log{(\mathbb{P}(\text{c}) / \mathbb{P}(\text{f}))}$ values for all 10 stars is evidence that these stars are more likely to belong to the \texttt{corespray} extra-tidal star distribution than the field.

When examining the locations of the highest-probability extra-tidal stars in Figure \ref{fig:final}, we see that the majority of stars with large $\log{(\mathbb{P}(\text{c}) / \mathbb{P}(\text{f}))}$ are the ones with highly negative radial velocities. This observation occurs despite the fact that these stars are not necessarily the ones that are located closest to M3's proper motion median. It should again be noted that $\log{(\mathbb{P}(\text{c}) / \mathbb{P}(\text{f}))}$ is computed from \textit{both} proper motion and radial velocity information. Thus, it is the \textit{combination of these three parameters} that determine the probabilities that an extra-tidal candidate is associated with each distribution. 

\begin{figure}
    \centering
    \includegraphics[width=\columnwidth]{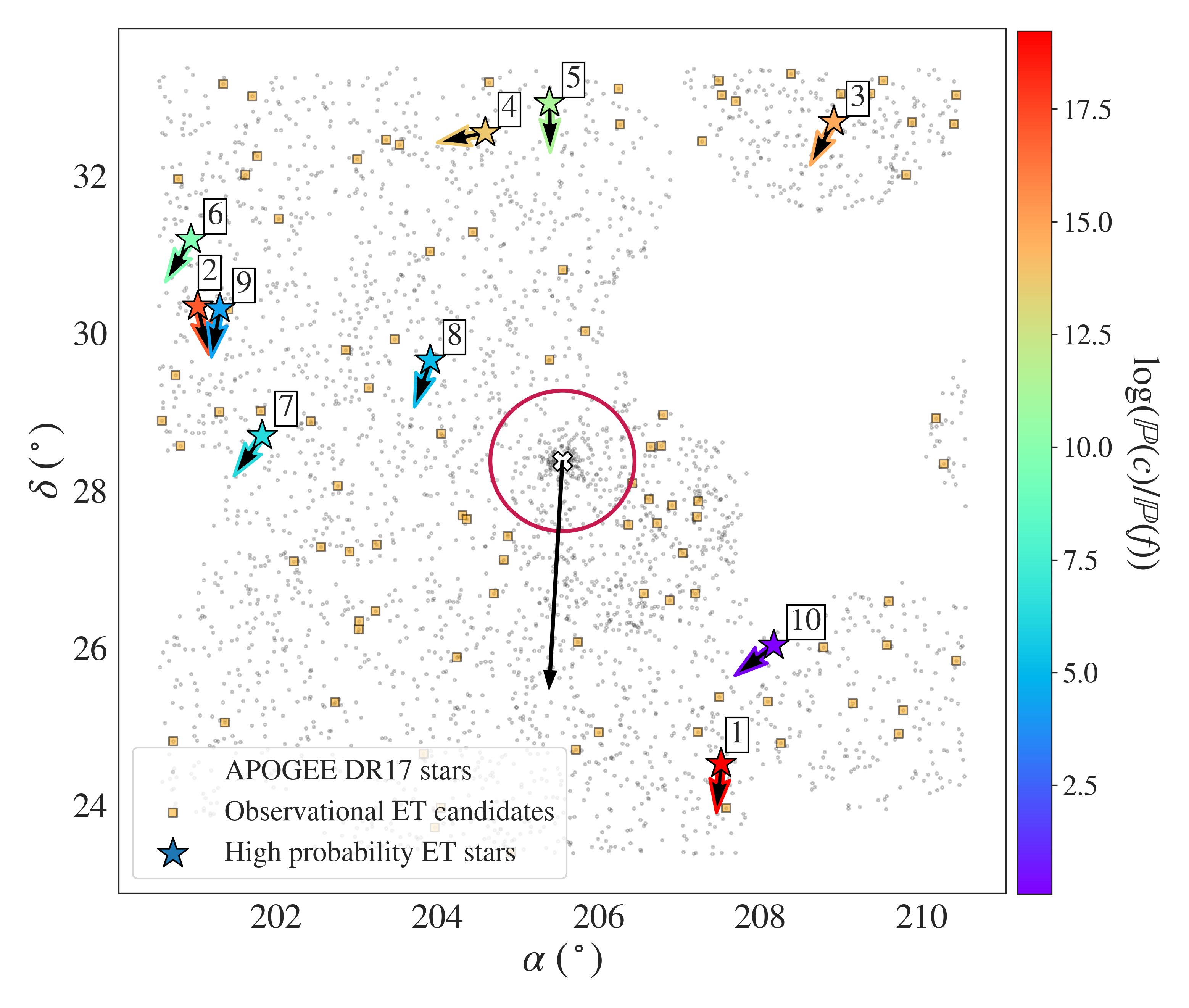}
    \caption{Spatial distribution of the 10 new high-probability extra-tidal stars of M3 identified in this study relative to APOGEE DR17 stars in a $10^{\circ} \times 10^{\circ}$ FOV around M3. Each extra-tidal star is numbered and coloured by its $\log{(\mathbb{P}(\text{c}) / \mathbb{P}(\text{f}))}$ value. Proper motion directions of each star are marked with coloured arrows. M3's tidal radius at apogalacticon of $r_{t} = 159.03$pc is shown as a magenta circle, its proper motion with a black arrow and the cluster centre with a white cross.}
    \label{fig:finalspatial}
\end{figure}

In Figure \ref{fig:final}, we see that the \texttt{corespray} radial velocity distribution is skewed to larger negative radial velocities than the field star distribution. Thus, stars that are located far from the radial velocity field star distribution (i.e. stars that have highly negative radial velocities) will be assigned larger probabilities of being associated with the \texttt{corespray} distribution, even though they are not necessarily "close" to M3's proper motion median. Hence, we observe the high $\log{(\mathbb{P}(\text{c}) / \mathbb{P}(\text{f}))}$ of these high-velocity stars. The spatial locations of our 10 new high-probability extra-tidal stars relative to APOGEE DR17 field stars in a $10^{\circ} \times 10^{\circ}$ FOV around M3 in are shown in Figure \ref{fig:finalspatial}.

\subsection{Alternative Probability Tests}

While odds ratio computations allow us to statistically infer the highest-probability extra-tidal candidates of M3, other kinematic and photometric tests could also be employed. Specifically, utilizing conserved kinematic quantities like actions would be the ideal way to probe extra-tidal association. Actions are useful because the actions of an extra-tidal star will remain similar to those of the parent cluster itself, no matter where along the cluster’s orbit the star escaped \citep{2008gady.book.....B}. 

Computing the actions of a star requires an assumed Galactic potential and the knowledge of six parameters: $\alpha$, $\delta$, $\mu_{\alpha}$, $\mu_{\delta}$, $v_{r}$ and distance ($d$). While the APOGEE DR17 actions of the high-probability extra-tidal candidates are consistent with the \texttt{corespray} distribution within error, uncertainties in all six components result in the errors of the actions being large. The most uncertain input parameter is the \texttt{astroNN} distance estimate, with most extra-tidal candidates having a $\delta d/d  > 0.20$. Alternative methods for measuring distances to Gaia EDR3 stars have improved fractional distance errors, with the extra-tidal candidates having a mean $\delta d/d \sim 0.09$ \citep{2021AJ....161..147B}. While using this catalogue would indeed allow many candidates to have smaller distance uncertainties, more than one-third of our extra-tidal candidates would still have $\delta d/d > 0.10$. Similarly, action computation of the field star distribution would also be challenging, as field stars span a wide range of distances. Since \cite{2021AJ....161..147B} state that their distance estimates are only reliable out to several kpc, and M3 is located 10.2 kpc away, we discard this method of action analysis until more accurate distances to faraway stars are obtained. Nevertheless, computation of the odds ratios using proper motions and radial velocities remains an acceptable alternative to probe the kinematic associations of extra-tidal stars to M3 (and other Galactic GCs). 

\subsection{Extra-Tidal Versus Tidal Tail Stars}

As previously mentioned, we have only considered three-body core encounters as the dynamical interaction that produces extra-tidal stars. However, other dynamical processes like tidal stripping can also cause stars to migrate beyond the tidal radius of a GC, albeit resulting in stars populating tidal tails rather than being isolated field stars. To determine if any of our high-probability extra-tidal stars are more likely to be associated with M3's tidal tails than with extra-tidal stars produced via three-body encounters (i.e. a \texttt{corespray} distribution) we first generate mock tidal tails for M3. The mock tidal tails are simulated using the particle spray method of \citet{Fardal2015}, which has recently been implemented in \texttt{galpy} \citep{2015ApJS..216...29B, Banik2019}. For the simulation, we generate 10,000 stars over a disruption timescale of 1 Gyr in order to ensure the tails are well populated in our FOV. The same structural and orbital parameters for M3 as above are assumed \citep{2018MNRAS.478.1520B, Vasiliev2019}. The cluster's orbit is integrated in the \texttt{MWPotential2014} Milky Way galaxy model from \citet{2015ApJS..216...29B} and the orbits of kicked stars are integrated in the combined potential of the Milky Way and the cluster itself. The cluster's potential is the same as the one used for the \texttt{corespray} simulation.

The kinematic distributions of the mock tail stars are not Gaussian within the FOV, making it not possible to repeat a similar analysis to the one presented in Section \ref{sec:confirmation}. In fact, the distributions are bimodal, as the stars escape the cluster with velocities slightly greater or less than the cluster itself. However the distributions themselves are quite narrow, with $\mu_{\alpha}$ and $\mu_{\delta}$ differing from M3 itself by at most 0.3 mas/yr and $v_{r}$ differing from M3 by at most 25 km/s. Such narrow distributions are expected given that stars escape with low velocities. Given that no candidate stars have kinematics so similar to M3 in Figure \ref{fig:logodds}, we can conclude that none of the candidates are consistent with being tail stars. The lack of any tidal tail stars about M3 is consistent with \cite{2010A&A...522A..71J} and \cite{2014MNRAS.445.2971C}. However, it should also be noted that even if some tail stars existed about M3, they would most likely be low mass main sequence stars below APOGEE's detection limit due to the effects of mass segregating \citep{Webb2022}.

\subsection{Additional Ejection Mechanisms and Influences}\label{sec:ejectionstuff}

While much discussion has been paid to the 10 new high-probability extra-tidal stars of M3, the additional 93 stars that are chemically similar to M3 should not necessarily be disregarded. In this study, we use \texttt{corespray} to simulate three-body interactions in M3's core. However, it is possible that the other chemically similar stars identified in the dimensionality reductions escaped M3 via different dynamical processes. For instance, \cite{leigh11} show a wide range of small-$N$ body interactions can occur within a dense star cluster, many of which leading to a star's ejection. Natal kicks \citep{2004ApJ...607L...9M} or ejections via supernovae \citep{2018ApJ...865...15S, 2022arXiv220914319K} could also cause stars to escape a cluster, all while altering their kinematics. Alternatively, some of these stars could have been produced via three-body encounters, albeit at earlier times in the past. Thus these stars could have experienced kinematic perturbations from the underlying Galactic tidal field, which is more complex than the tidal field assumed in this study, causing their present day kinematic quantities to differ from the motions they had when they left the cluster.

To probe whether any of the additional 93 candidate stars could have been ejected from the core of M3, albeit via a different interaction than considered here, we perform an additional \texttt{corespray} simulation where kick velocities are drawn from a uniform distribution between M3's core escape velocity and 500 km/s. We find that the kinematic distribution of extra-tidal stars in the uniform kick velocity distribution is much wider compared to the the three-body kick velocity distribution (see Figure \ref{fig:uniform} in the Appendix). We can't use the uniform kick velocity kinematic distribution to calculate probabilities like we did with the three-body kick kinematic distribution because it is not dynamically motivated. However, we can qualitatively find that approximately half of the candidate stars fall within the kinematic parameter space that \texttt{corespray} stars generated from the uniform kick velocity distribution. These are stars that \textit{could} have escaped M3 given a high enough velocity kick, but the mechanism that leads to their escape is not constrained by this study.

\section{Conclusions}\label{sec:conclusions}

In this study, we present a new methodology that combines observational and theoretical techniques to identify 10 new extra-tidal stars of the Galactic GC M3. Using two different unsupervised machine learning algorithms (t-SNE and UMAP), we identify stars beyond the tidal radius of M3 that are chemically similar to a control group of cluster members of M3. While machine learning clustering algorithms like t-SNE and UMAP are excellent tools in identifying stars with similar chemical abundances, particle-spray simulations like \texttt{corespray} are necessary for tracing the extra-tidal candidates back to their suspected birth cluster and assigning probabilities of association. Ultimately, each of our identified extra-tidal stars has passed rigorous tests to confirm its extra-tidal nature. The results and implications can be summarized as follows:

\begin{enumerate}

    \item An application of the t-SNE and UMAP dimensionality-reduction algorithms to stars within a $10^{\circ} \times 10^{\circ}$ FOV around M3 identifies 103 extra-tidal candidates that are chemically similar to a control group of stars within M3 (Figure \ref{fig:tsne}). Chemical abundance distributions further confirm that the 103 observationally-identified extra-tidal candidates have similar abundances to M3; both samples occupying a unique chemical location relative to stars in the full FOV (Figure \ref{fig:violin}).
    
    \item A \texttt{corespray} simulation of 40,000 extra-tidal stars of M3 (Figure \ref{fig:cspray}) finds that only 3394 stars are located within a $10^{\circ} \times 10^{\circ}$ FOV around M3 (our observational FOV). This result indicates that the majority of core three-body interactions that could have occurred over M3's past orbital period result in extra-tidal stars being kicked beyond this FOV. Hence, any extra-tidal stars found within the FOV will have either escaped M3 recently or with a low kick-velocity.
    
    \item A multivariate Gaussian model and an XD compute the probabilities that each extra-tidal candidate either belongs to a \texttt{corespray} distribution of extra-tidal stars of M3 or the field stars around M3 (Figure \ref{fig:logodds}). By computing $\log{(\mathbb{P}(\text{c}) / \mathbb{P}(\text{s}))}$, we find that 10 stars have $\log{(\mathbb{P}(\text{c}) / \mathbb{P}(\text{s}))} > 0$  (Table \ref{tab:prob}). This result indicates that although stars beyond the tidal radius of a GC can be chemically similar, only a fraction have kinematics that are consistent with being extra-tidal.  %\\ 
    
    \item A proper motion and radial velocity analysis highlights that while all 10 extra-tidal candidates have properties consistent with M3, some were likely produced in rare three-body interactions (Figure \ref{fig:final}). This finding suggests that imposing proper motion and radial velocity constraints when using high-dimensional analysis to initially search for extra-tidal stars could result in high-velocity extra-tidal stars being missed.

    \item A meaningful action variable analysis is unable to be performed due to large uncertainties in distances and pre-computed APOGEE DR17 actions. Once more accurate distances to stars at $\sim 10$\ kpc are obtained, conserved kinematic quantities like actions can provide additional constraints in defining high-probability extra-tidal stars of M3 and other Galactic GCs at similar distances.

    \item None of the extra-tidal candidate stars are consistent with being members of M3's tidal tails. However, we find that approximately half of the candidate stars could be extra-tidal stars of M3 if they were given sufficient velocity kicks due to a mechanism that is not a three-body interaction. 
\end{enumerate}

Ultimately, all 10 extra-tidal stars identified in this study have passed spatial, chemical and kinematic analyses, providing strong evidence that each candidate is indeed an extra-tidal star of M3. None of the eight extra-tidal stars presented in \cite{2016ApJ...829..123N} are recovered in this study, as (i) APOGEE and LAMOST observe different regions of the sky (eliminating seven out of \cite{2016ApJ...829..123N}  eight stars) and (ii) the proper motion of the remaining extra-tidal star is far different than the stars in our \texttt{corespray} sample. Thus, \textit{each extra-tidal star presented in this work is a new extra-tidal star of M3}. As new extra-tidal stars are discovered, one will be able to better understand core dynamics and star formation histories in GCs. Furthermore, as \texttt{corespray} also computes the orbital parameters of the recoiled binaries of the simulated three-body systems, binary fractions and their locations in the Galaxy can also be inferred. Thus, future applications of \texttt{corespray} to other Galactic GCs have the potential to inform us not just about GC evolution, but formation and evolution of our Galaxy itself.

\section*{Acknowledgements}
 The authors wish to thank Samantha Berek, Jo Bovy, Yuyang Chen, Bolin Fan, Henry Leung, Ting Li, Ayush Pandhi, Dang Pham, Tristan Wheeler and the anonymous referee for valuable feedback, conversations and suggestions for improvement throughout the duration of this study and the preparation of this manuscript. NWCL gratefully acknowledges the generous support of a Fondecyt Iniciaci\'on grant 11180005, as well as support from Millenium Nucleus NCN19-058 (TITANs) and funding via the BASAL Centro de Excelencia en Astrofisica y Tecnologias Afines (CATA) grant PFB-06/2007.  NWCL also thanks support from ANID BASAL project ACE210002 and ANID BASAL projects ACE210002 and FB210003.

\section*{Data Availability}
All data and parameters in this study are obtained using \texttt{apogee tools} \citep{2016ApJ...817...49B}, which can be downloaded at \url{https://github.com/jobovy/apogee}. Chemical abundances are acquired from the \texttt{astroNN} catalogues \citep{2019MNRAS.483.3255L, 2019MNRAS.489.2079L}, which can be downloaded at \url{https://astronn.readthedocs.io/en/latest/}. The \texttt{corespray} simulation software is available for download at \url{https://github.com/webbjj/corespray}. Important individual GC structural parameters of M3 are obtained from \cite{2018MNRAS.478.1520B}, where the online database is accessible at \url{https://people.smp.uq.edu.au/HolgerBaumgardt/globular/}. All high-dimensional analysis and multivariate Gaussian modelling are performed using \texttt{scikit-learn} \citep{scikit-learn}, which can be accessed at \url{https://scikit-learn.org/stable/}. Finally, the XD technique is implemented via the \texttt{astroML} software and can be installed from \url{https://www.astroml.org/}.



\bibliographystyle{mnras}
\bibliography{example} % if your bibtex file is called example.bib

%\begin{thebibliography}{99}
%\end{thebibliography}

%%%%%%%%%%%%%%%%%%%%%%%%%%%%%%%%%%%%%%%%%%%%%%%%%%

%%%%%%%%%%%%%%%%% APPENDICES %%%%%%%%%%%%%%%%%%%%%

\appendix

\section{Extra-Tidal Candidate Parameters} \label{sec:appendix}

As mentioned in Section \ref{sec:observational}, we identify 103 stars located beyond M3's tidal radius that have similar chemical abundances to a control group of M3 members. In Section \ref{sec:newstars}, we observe that 10 of these observationally-identified extra-tidal candidates have higher probabilities of belonging to a simulated \texttt{corespray} distribution of M3 extra-tidal stars than the surrounding field stars. Although the other 93 extra-tidal candidates have higher probabilities of being associated with the field rather than with \texttt{corespray}, these stars still exhibit chemical similarities to M3 members (Figure \ref{fig:tsne}). As described in Section \ref{sec:ejectionstuff}, it is possible that these candidates could still have a connection to M3, albeit being produced via a different dynamical interaction or having been perturbed by the Milky Way's tidal field. Thus, in Table \ref{tab:allstars} we list the APOGEE DR17 identifiers, spatial locations, proper motions, radial velocities and probabilities of belonging to the \texttt{corespray} extra-tidal distribution for all 103 observationally-identified extra-tidal candidates. The 10 high-probability extra-tidal stars are indicated with a $\star$.

\onecolumn
\begin{center}
\setlength\tabcolsep{12pt}
\begin{longtabu}{|l|l|l|l|l|l|l|l|l|l|l|}
\captionsetup{width=17.5cm}
\caption{Spatial, kinematic and probability parameters of the full 103 observationally-identified extra-tidal star sample around M3. While some of these stars had lower probabilities of belonging to the \texttt{corespray} extra-tidal star distribution, they are chemically similar to a control group of cluster members of M3. Parameters obtained from the APOGEE DR17 catalogue \citep{2022ApJS..259...35A} are marked with a $^{*}$, parameters obtained from \texttt{astroNN} are marked with a $^{\wr}$ \citep{2019MNRAS.483.3255L, 2019MNRAS.489.2079L, 2019MNRAS.490.4740B} and parameters obtained from Gaia EDR3 \citep{2021A&A...649A...1G} are marked with a $^{\dagger}$. The 10 high-probability extra-tidal stars are indicated with a $\star$.}
  \label{tab:allstars} \\
\hline \multicolumn{1}{|c|}{APOGEE ID$^{*}$} & \multicolumn{1}{c|}{$\alpha^{*}$} & \multicolumn{1}{c|}{$\delta^{*}$} & \multicolumn{1}{c|}{$d^{\wr} $}   & \multicolumn{1}{c|}{$\mu_{\alpha}^{\dagger}$}  & \multicolumn{1}{c|}{$\mu_{\delta}^{\dagger}$}  & \multicolumn{1}{c|}{${v_{r}}^{*}$}  & \multicolumn{1}{c|}{$\log{(\mathbb{P}(\text{c}) / \mathbb{P}(\text{f}))}$} \\ 

\multicolumn{1}{|c|}{} & \multicolumn{1}{c|}{[deg]} & \multicolumn{1}{c|}{[deg]}   &\multicolumn{1}{c|}{[pc]} & \multicolumn{1}{c|}{[mas/yr]}  & \multicolumn{1}{c|}{[mas/yr]}  & \multicolumn{1}{c|}{[km/s]} & \multicolumn{1}{c|}{} \\ \hline 

\endfirsthead
\multicolumn{8}{c}%

{{\bfseries \tablename\ \thetable{} -- continued from previous page.}} \\
\hline \multicolumn{1}{|c|}{APOGEE ID$^{*}$} & \multicolumn{1}{c|}{$\alpha^{*}$} & \multicolumn{1}{c|}{$\delta^{*}$} & \multicolumn{1}{c|}{$d^{\wr} $}   & \multicolumn{1}{c|}{$\mu_{\alpha}^{\dagger}$}  & \multicolumn{1}{c|}{$\mu_{\delta}^{\dagger}$}  & \multicolumn{1}{c|}{${v_{r}}^{*}$}  & \multicolumn{1}{c|}{$\log{(\mathbb{P}(\text{c}) / \mathbb{P}(\text{f}))}$} \\ 

\multicolumn{1}{|c|}{} & \multicolumn{1}{c|}{[deg]} & \multicolumn{1}{c|}{[deg]} &\multicolumn{1}{c|}{[pc]}   & \multicolumn{1}{c|}{[mas/yr]}  & \multicolumn{1}{c|}{[mas/yr]}  & \multicolumn{1}{c|}{[km/s]} & \multicolumn{1}{c|}{}  \\ \hline 
\endhead

\hline \multicolumn{8}{|r|}{{Continued on next page.}} \\ \hline
\endfoot
\hline
\endlastfoot
2M13221962+2853203 & 200.581 & 28.888 & 418.45 & -233.614 & -286.569 & -55.512 & -178695.482 \\
2M13225389+2448532 & 200.724 & 24.814 & 9289.86 & -2.094 & -3.826 & -43.63 & -3.413  \\
2M13230059+2927561 & 200.752 & 29.465 & 2200.571 & -26.797 & -14.122 & 83.181 & -1301.593  \\
2M13230915+3157296 & 200.788 & 31.958 & 3687.256 & -9.291 & -14.912 & -61.228 & -304.625  \\
2M13231525+2834137 & 200.813 & 28.57 & 464.202 & -239.481 & -44.355 & -50.988 & -89765.434  \\
$\star$ 2M13234701+3111279 & 200.945 & 31.191 & 13846.063 & -1.756 & -2.846 & -116.314 & 9.749  \\
$\star$ 2M13240682+3020316 & 201.028 & 30.342 & 4421.892 & 1.272 & -5.464 & -189.679 & 17.002  \\
2M13251080+2900003 & 201.295 & 29.0 & 5748.236 & -9.713 & -4.014 & 3.161 & -137.965  \\
$\star$ 2M13251237+3018535 & 201.301 & 30.314 & 7344.123 & -1.13 & -6.438 & -160.858 & 4.307  \\
2M13252308+3310108 & 201.346 & 33.169 & 22747.718 & -1.757 & -1.612 & 54.494 & -5.029  \\
2M13252729+2503126 & 201.363 & 25.053 & 3046.792 & -9.326 & 2.223 & -18.561 & -109.76  \\
2M13253774+3018220 & 201.407 & 30.306 & 4835.265 & -4.207 & -9.506 & -42.262 & -79.259  \\
2M13262865+3200458 & 201.619 & 32.012 & 7762.88 & -5.355 & -3.065 & 14.557 & -39.298  \\
2M13264876+3300388 & 201.703 & 33.01 & 7246.469 & -2.371 & -5.087 & -110.683 & -2.751  \\
2M13270393+3215104 & 201.766 & 32.252 & 2112.221 & 4.622 & -16.842 & -149.049 & -160.181  \\
2M13271442+2900409 & 201.81 & 29.011 & 1428.097 & -31.06 & -47.74 & -41.122 & -3807.367  \\
$\star$ 2M13271850+2841521 & 201.827 & 28.697 & 18603.378 & -1.568 & -2.231 & -73.028 & 6.358  \\
2M13280718+3127198 & 202.029 & 31.455 & 5923.512 & 0.816 & -10.302 & -99.347 & -42.124  \\
2M13285221+2705512 & 202.217 & 27.097 & 5293.979 & -3.165 & -6.959 & -173.036 & -13.418  \\
2M13294218+2852493 & 202.425 & 28.88 & 3018.643 & -42.454 & -26.872 & 156.198 & -3618.737  \\
2M13301333+2717046 & 202.555 & 27.284 & 4156.3 & -0.712 & -18.958 & -178.933 & -207.293  \\
2M13305572+2518385 & 202.732 & 25.31 & 3974.775 & -7.637 & -10.19 & -58.625 & -155.059  \\
2M13310270+2803397 & 202.761 & 28.061 & 6704.045 & -2.78 & -4.369 & -64.231 & -8.264  \\
2M13312657+2947187 & 202.86 & 29.788 & 739.22 & 15.237 & -113.417 & -6.728 & -10410.135  \\
2M13313781+2713289 & 202.907 & 27.224 & 3109.405 & 4.994 & -11.641 & -13.592 & -104.377  \\
2M13320138+3212483 & 203.005 & 32.213 & 529.207 & -97.777 & -80.343 & -104.566 & -22511.237  \\
2M13320566+2614003 & 203.023 & 26.233 & 9871.155 & -2.731 & -4.913 & -137.845 & -0.781  \\
2M13320644+2620099 & 203.026 & 26.336 & 4620.158 & -9.442 & -9.242 & -112.258 & -180.055  \\
2M13323523+2918295 & 203.146 & 29.308 & 3175.738 & -1.949 & -17.01 & -76.786 & -197.821  \\
2M13325603+2627593 & 203.233 & 26.466 & 7557.041 & -2.558 & -7.847 & -149.407 & -19.722  \\
2M13325885+2718504 & 203.245 & 27.314 & 6231.899 & -10.174 & -2.254 & 122.425 & -143.61  \\
2M13332728+3227326 & 203.363 & 32.459 & 2759.147 & -13.509 & -13.077 & 176.94 & -453.152  \\
2M13335179+2955261 & 203.465 & 29.923 & 14820.654 & -4.484 & -1.695 & 36.146 & -24.294  \\
2M13340645+3223497 & 203.526 & 32.397 & 2684.888 & -2.661 & -13.385 & 66.794 & -152.393  \\
2M13351884+2439111 & 203.828 & 24.653 & 163.775 & -189.121 & 31.3 & -52.838 & -50466.989  \\
2M13353738+3102263 & 203.905 & 31.04 & 541.548 & -87.566 & -53.618 & 75.996 & -15459.45  \\
$\star$ 2M13353852+2939287 & 203.91 & 29.657 & 6552.756 & -2.236 & -6.434 & -205.337 & 5.118  \\
2M13355149+2343075 & 203.964 & 23.718 & 1262.867 & -9.426 & -1.502 & 47.752 & -116.71  \\  
2M13360924+2358096 & 204.038 & 23.969 & 5923.698 & -16.814 & 0.602 & 135.273 & -373.212  \\
2M13361042+2843314 & 204.043 & 28.725 & 540.263 & -82.954 & -15.896 & 51.014 & -10652.548  \\
2M13365674+2553016 & 204.236 & 25.883 & 2184.551 & -25.688 & -22.565 & 63.794 & -1554.618  \\
2M13371399+2741118 & 204.308 & 27.686 & 2691.399 & 1.504 & -5.265 & -43.34 & -7.062  \\
2M13372695+2738074 & 204.362 & 27.635 & 2910.639 & -28.813 & -0.395 & -32.718 & -1161.344  \\
2M13374538+3117169 & 204.439 & 31.288 & 13732.833 & -3.529 & -1.224 & 121.074 & -18.564  \\
$\star$ 2M13382215+3233031 & 204.592 & 32.55 & 7675.512 & -2.184 & -0.45 & -191.682 & 13.763  \\
2M13383329+3311212 & 204.638 & 33.189 & 2328.157 & -1.281 & -29.694 & -116.427 & -645.044  \\
2M13384718+2641264 & 204.696 & 26.69 & 8351.948 & -1.926 & -8.614 & -97.11 & -32.303  \\
2M13391725+2707122 & 204.821 & 27.12 & 1083.342 & -33.237 & -12.553 & -80.604 & -1823.959  \\
2M13392970+2725086 & 204.873 & 27.419 & 1347.444 & -18.862 & -13.228 & -37.422 & -698.49  \\
2M13393858+2324093 & 204.91 & 23.402 & 2803.387 & -18.843 & -9.022 & 244.367 & -616.473  \\
2M13413310+2939316 & 205.387 & 29.658 & 1253.692 & -48.445 & -10.534 & -91.72 & -3626.921  \\
$\star$ 2M13413296+3255410 & 205.387 & 32.928 & 5910.173 & 0.096 & -8.231 & -244.763 & 11.305  \\
2M13421226+3048148 & 205.551 & 30.804 & 1940.452 & -31.733 & -3.141 & -96.47 & -1454.081  \\
2M13425125+2442197 & 205.713 & 24.705 & 2196.549 & -15.501 & -8.553 & 60.006 & -422.248  \\
2M13425643+2604283 & 205.735 & 26.074 & 10111.364 & -4.662 & -2.412 & 24.875 & -27.849  \\
2M13431999+3001209 & 205.833 & 30.022 & 11764.436 & -1.595 & -2.779 & 136.178 & -13.722  \\
2M13435960+2455333 & 205.998 & 24.925 & 570.664 & 7.577 & -8.218 & 3.71 & -110.772  \\
2M13445801+3306304 & 206.241 & 33.108 & 8645.228 & -3.811 & -3.697 & 29.321 & -24.423  \\
2M13450302+3239166 & 206.262 & 32.654 & 2987.031 & -53.007 & 0.03 & 14.196 & -4003.035  \\
2M13452795+2734001 & 206.366 & 27.566 & 47296.464 & -1.144 & -0.754 & 169.045 & -9.429  \\
2M13453968+2805362 & 206.415 & 28.093 & 4972.64 & -6.902 & -2.083 & 144.143 & -67.201  \\
2M13461283+2641252 & 206.553 & 26.69 & 4600.086 & -1.702 & -13.81 & 111.228 & -161.939  \\
2M13462942+2753266 & 206.622 & 27.89 & 3088.608 & -2.155 & -15.265 & -51.806 & -162.629  \\
2M13463425+2833348 & 206.642 & 28.559 & 4212.972 & -0.906 & -13.501 & -205.726 & -73.784  \\
2M13465357+2735089 & 206.723 & 27.585 & 2471.714 & -50.672 & -26.497 & -132.811 & -4745.083  \\
2M13470503+2834136 & 206.77 & 28.57 & 16598.224 & -1.593 & -1.209 & 20.833 & -1.695  \\
2M13471172+2857471 & 206.798 & 28.963 & 4743.041 & -14.962 & -7.277 & -90.816 & -360.342  \\
2M13473000+2636130 & 206.875 & 26.603 & 2263.907 & -7.89 & -17.271 & -62.865 & -327.737  \\
2M13473772+2748479 & 206.907 & 27.813 & 8267.7 & -5.094 & -3.829 & 12.701 & -39.55  \\
2M13480854+2712278 & 207.035 & 27.207 & 6036.497 & 0.599 & -9.622 & -144.539 & -23.821  \\
2M13484600+2641250 & 207.191 & 26.69 & 3942.425 & -1.542 & -11.163 & -106.196 & -62.71  \\
2M13485213+2740093 & 207.217 & 27.669 & 224.053 & -234.975 & -198.961 & -50.308 & -134386.405  \\
2M13485492+2455467 & 207.228 & 24.929 & 4621.769 & -6.915 & -17.607 & -174.538 & -280.543  \\
2M13490745+3226207 & 207.281 & 32.439 & 1501.892 & -0.053 & -29.16 & -71.615 & -623.469  \\
2M13485603+2751541 & 207.233 & 27.865 & 7080.795 & -8.147 & -1.59 & 125.37 & -89.214  \\
2M13495713+3312327 & 207.488 & 33.209 & 4057.047 & -2.376 & -9.061 & -29.203 & -53.874  \\
2M13495806+2522400 & 207.491 & 25.377 & 5158.221 & -5.777 & 0.234 & -20.387 & -36.666  \\
$\star$ 2M13500350+2431542 & 207.514 & 24.531 & 9663.527 & -0.541 & -5.356 & -199.733 & 19.233  \\
2M13500435+3301440 & 207.518 & 33.028 & 5588.955 & -4.943 & -9.793 & -48.061 & -94.202  \\
2M13500482+2427061 & 207.52 & 24.451 & 4572.004 & -9.749 & -2.686 & 68.862 & -132.846  \\
2M13501865+2357507 & 207.577 & 23.964 & 19619.564 & -2.658 & -1.124 & 136.996 & -13.182  \\
2M13504647+3256579 & 207.693 & 32.949 & 1659.718 & -24.038 & -0.745 & -141.824 & -804.864  \\
2M13522181+2519046 & 208.09 & 25.317 & 608.27 & -45.196 & -56.555 & 136.265 & -6685.506  \\
$\star$ 2M13524016+2601592 & 208.167 & 26.033 & 8461.93 & -3.425 & -2.624 & -132.903 & 0.081  \\
2M13530057+2447273 & 208.252 & 24.79 & 2455.072 & -37.238 & -5.82 & 40.441 & -2059.164  \\
2M13533112+3317545 & 208.379 & 33.298 & 5718.64 & -7.795 & -7.435 & 25.513 & -128.22  \\
2M13550799+2600242 & 208.783 & 26.006 & 10707.711 & -0.407 & -3.037 & 33.309 & -3.883  \\
$\star$ 2M13553890+3241208 & 208.912 & 32.689 & 16030.677 & -1.268 & -2.299 & -138.898 & 14.809  \\
2M13555858+3302315 & 208.994 & 33.042 & 5813.009 & -6.936 & -1.693 & 96.118 & -64.251  \\
2M13563485+2517471 & 209.145 & 25.296 & 12684.391 & -2.773 & -3.067 & -67.637 & -2.576  \\
2M13572764+3302442 & 209.365 & 33.045 & 4588.824 & -11.484 & -5.632 & 14.797 & -210.371  \\
2M13580623+3312372 & 209.525 & 33.21 & 10350.6 & -6.794 & -1.329 & 161.259 & -62.305  \\
2M13581572+2602122 & 209.565 & 26.036 & 1769.465 & -21.404 & -17.072 & -85.888 & -972.231  \\
2M13582241+2635398 & 209.593 & 26.594 & 5644.837 & -4.761 & -3.952 & -12.575 & -33.226  \\
2M13585105+2454427 & 209.712 & 24.911 & 2140.677 & -29.907 & 1.325 & -138.452 & -1245.059  \\
2M13590531+2512219 & 209.772 & 25.206 & 5616.199 & 0.067 & -12.109 & -72.584 & -77.123  \\
2M13591443+3200423 & 209.81 & 32.011 & 6563.059 & -5.017 & -3.585 & -68.815 & -29.051  \\
2M13593064+3241036 & 209.877 & 32.684 & 4682.448 & -19.575 & -10.683 & -16.317 & -678.081  \\
2M14004220+2855119 & 210.175 & 28.919 & 3637.32 & -5.869 & -7.203 & -198.696 & -51.777  \\
2M14010561+2820306 & 210.273 & 28.341 & 8799.474 & -13.822 & -4.638 & -27.22 & -282.634  \\
2M14013546+3239245 & 210.397 & 32.656 & 1108.471 & -41.541 & -0.091 & -80.28 & -2451.459  \\
2M14014321+2550085 & 210.43 & 25.835 & 5852.216 & -14.037 & -1.366 & 8.016 & -265.724  \\
2M14014338+3301395 & 210.43 & 33.027 & 1647.218 & -46.407 & 27.245 & -77.696 & -3263.855  \\
\end{longtabu}
\end{center}

\begin{figure*}
    \centering
    \includegraphics[width=\textwidth]{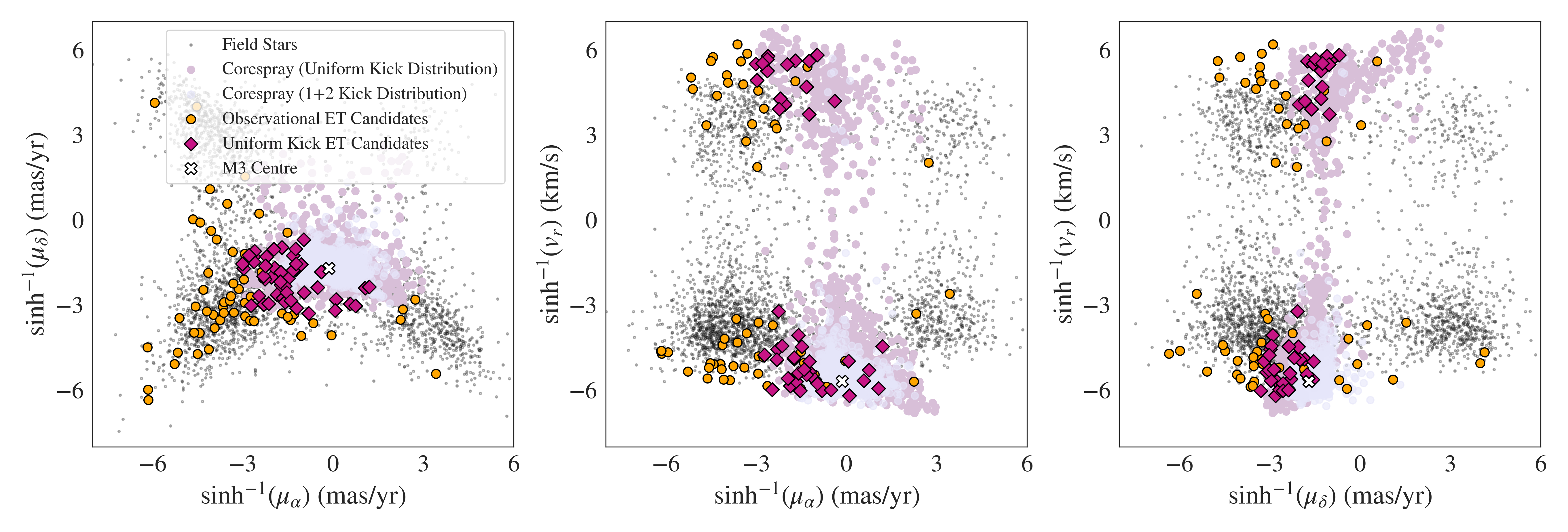}
    \caption{Similar to Figure \ref{fig:logodds} except now comparing \texttt{corespray} stars ejected via a uniform kick velocity distribution (pink circles) to \texttt{corespray} stars ejected via a 1+2 interaction (lavender circles). For comparison purposes, the 2998 field stars (black points) and 103 extra-tidal candidates (orange circles and magenta diamonds) are also shown. A uniform kick velocity distribution widens the \texttt{corespray} kinematic distribution compared to the \texttt{corespray} stars ejected via a 1+2 interaction. Approximately half of the extra-tidal candidate stars fall within this widened distribution, which are marked as magenta diamonds. Hence some of our additional extra-tidal candidates could be stars that escaped M3 via a mechanism other than a three body interaction. Each representation depicts the $\sinh^{-1}$ of the data, as it best separates low and high values to allow for optimal data visualization. M3's proper motion of $(\mu_{\alpha}, \mu_{\delta}) = (-0.142, -2.647) \ \text{mas/yr}$ and radial velocity of $v_{r} = -147.28 \ \text{km/s}$ are obtained from \citet{Vasiliev2019} and used to indicate the centre of M3 (white cross) in each panel.}
    \label{fig:uniform}
\end{figure*}

%%%%%%%%%%%%%%%%%%%%%%%%%%%%%%%%%%%%%%%%%%%%%%%%%%

\vspace{-1cm}
% Don't change these lines
\bsp	% typesetting comment
\label{lastpage}
\end{document}